\documentclass[citeautoscript, superscriptaddress,twocolumn,showpacs,prb,aps,10pt]{revtex4}

\usepackage{graphicx} \usepackage{amsmath, amssymb, bm}
\usepackage{dcolumn} \usepackage{color}
\usepackage{color}
\usepackage{eso-pic}
\usepackage{fix-cm}

\usepackage{multirow}
\usepackage{sidecap}

\usepackage{marginnote}
\usepackage{ulem} 

\renewcommand{\emph}[1]{\textit{#1}}

\begin{document}

\title{Anisotropic Contrast Optical Microscope}
\author{D.~Peev}
\affiliation{Department of Electrical and Computer Engineering, University of Nebraska-Lincoln, NE 68588, U.S.A.}
\affiliation{Center for Nanohybrid Functional Materials, University of Nebraska-Lincoln, NE 68588, U.S.A.}
\author{T.~Hofmann}
\affiliation{Department of Electrical and Computer Engineering, University of Nebraska-Lincoln, NE 68588, U.S.A.}
\affiliation{Center for Nanohybrid Functional Materials, University of Nebraska-Lincoln, NE 68588, U.S.A.}
\affiliation{Department of Physics, Chemistry, and Biology (IFM), Link{\"o}ping University, SE 581 83 Link{\"o}ping, Sweden}
\author{N.~Kananizadeh}
\affiliation{Department of Civil Engineering, University of Nebraska-Lincoln, NE 68588, U.S.A.}
\affiliation{Center for Nanohybrid Functional Materials, University of Nebraska-Lincoln, NE 68588, U.S.A.}
\author{S.~Beeram}
\affiliation{Department of Chemistry, University of Nebraska-Lincoln, NE 68588, U.S.A.}
\affiliation{Center for Nanohybrid Functional Materials, University of Nebraska-Lincoln, NE 68588, U.S.A.}
\author{E.~Rodriguez}
\affiliation{Department of Chemistry, University of Nebraska-Lincoln, NE 68588, U.S.A.}
\affiliation{Center for Nanohybrid Functional Materials, University of Nebraska-Lincoln, NE 68588, U.S.A.}
\author{S.~Wimer}
\affiliation{Department of Electrical and Computer Engineering, University of Nebraska-Lincoln, NE 68588, U.S.A.}
\affiliation{Center for Nanohybrid Functional Materials, University of Nebraska-Lincoln, NE 68588, U.S.A.}
\author{K.~B.~Rodenhausen}
\affiliation{Biolin Scientific Inc., Paramus, NJ 07652, U.S.A.}
\author{C.~M.~Herzinger}
\affiliation{J.~A.~Woollam~Co., Inc., Lincoln, Nebraska 68508-2243, U.S.A.}
\author{T.~Kasputis}
\affiliation{Department of Biomedical Engineering, University of Michigan, Ann Arbor, MI 48109, USA}
\author{E.~Pfaunmiller}
\affiliation{Celerion, Inc., Lincoln, NE 68502, USA}
\author{A.~Nguyen}
\affiliation{Department of Biological Systems Engineering, University of Nebraska-Lincoln, NE 68583, USA}
\author{R.~Korlacki}
\affiliation{Department of Electrical and Computer Engineering, University of Nebraska-Lincoln, NE 68588, U.S.A.}
\affiliation{Center for Nanohybrid Functional Materials, University of Nebraska-Lincoln, NE 68588, U.S.A.}
\author{A.~Pannier}
\affiliation{Department of Biological Systems Engineering, University of Nebraska-Lincoln, NE 68583, USA}
\affiliation{Center for Nanohybrid Functional Materials, University of Nebraska-Lincoln, NE 68588, U.S.A.}
\author{Y.~Li}
\affiliation{Department of Civil Engineering, University of Nebraska-Lincoln, NE 68588, U.S.A.}
\affiliation{Center for Nanohybrid Functional Materials, University of Nebraska-Lincoln, NE 68588, U.S.A.}
\author{E.~Schubert}
\affiliation{Department of Electrical and Computer Engineering, University of Nebraska-Lincoln, NE 68588, U.S.A.}
\affiliation{Center for Nanohybrid Functional Materials, University of Nebraska-Lincoln, NE 68588, U.S.A.}
\author{D.~Hage}
\affiliation{Department of Chemistry, University of Nebraska-Lincoln, NE 68588, U.S.A.}
\affiliation{Center for Nanohybrid Functional Materials, University of Nebraska-Lincoln, NE 68588, U.S.A.}
\author{M.~Schubert}
\email{schubert@engr.unl.edu}\homepage{http://ellipsometry.unl.edu}
\affiliation{Department of Electrical and Computer Engineering, University of Nebraska-Lincoln, NE 68588, U.S.A.}
\affiliation{Center for Nanohybrid Functional Materials, University of Nebraska-Lincoln, NE 68588, U.S.A.}
\affiliation{Department of Physics, Chemistry, and Biology (IFM), Link{\"o}ping University, SE 581 83 Link{\"o}ping, Sweden}
\affiliation{Leibniz Institute of Polymer Research (IPF) Dresden, D 01005 Dresden, Germany}

\begin{abstract}
An optical microscope is described that reveals contrast in the Mueller matrix images of a thin, transparent or semi-transparent specimen located within an anisotropic object plane (anisotropic filter). The specimen changes the anisotropy of the filter and thereby produces contrast within the Mueller matrix images. Here we use an anisotropic filter composed of a semi-transparent, nanostructured thin film with sub-wavelength thickness placed within the object plane. The sample is illuminated as in common optical microscopy but the light is modulated in its polarization using combinations of linear polarizers and phase plate (compensator) to control and analyze the state of polarization. Direct generalized ellipsometry data analysis approaches permit extraction of fundamental Mueller matrix object plane images dispensing with the need of Fourier expansion methods. Generalized ellipsometry model approaches are used for quantitative image analyses. These images are obtained from sets of multiple images obtained under various polarizer, analyzer, and compensator settings. Up to 16 independent Mueller matrix images can be obtained, while our current setup is limited to 11 images normalized by the unpolarized intensity. We demonstrate the anisotropic contrast optical microscope by measuring lithographically defined micro-patterned anisotropic filters, and we quantify the adsorption of an organic self-assembled monolayer film onto the anisotropic filter. Comparison with an isotropic glass slide demonstrates the image enhancement obtained by our method over microscopy without use of an anisotropic filter. In our current instrument we estimate the limit of detection for organic volumetric mass within the object plane of $\approx$ 49 fg within $\approx$ 7$\times$7~$\mu$m$^2$ object surface area. Compared to a quartz crystal microbalance with dissipation instrumentation, where contemporary limits require a total load of $\approx$ 500~pg for detection, the instrumentation demonstrated here improves sensitivity to a total mass required for detection by 4 orders of magnitude. We detail design and operation principles of the anisotropic contrast optical microscope, and we present further applications to detection of nanoparticles, to novel approaches for imaging chromatography, and to new contrast modalities for observations on living cells.
\end{abstract}


\maketitle

\section{Introduction}

Visible light optical microscopy methods are used to obtain magnified images of small samples. Advanced instruments include improved scanning (e.g., confocal microscopes) and detection (e.g., charge-coupled detector arrays) modes. Modulation-contrast enhancements such as polarization (petrographic) and phase contrast (Zernike) methods provide a fundamentally different source of contrast for specimens that are often entirely transparent for standard bright-field microscopy. In phase contrast microscopy phase shifts of light passing through a transparent specimen are converted to brightness changes in the image.\cite{Zernike1955} The contrast enhancement is obtained by emphasizing the phase changes against the phase of the isotropic background. In Nomarski interference contrast microscopy (NIC), also known as differential interference contrast (DIC) microscopy a polarized light source is separated into two orthogonally polarized, mutually coherent parts.\cite{NomarskiJPR1955,NomarskiRM1955} The two polarized components interact with the sample under a shear angle, and recombine before observation. The information contained within the two-beam interference is sensitive to polarization rotation caused by birefringence or optical activity within the specimen. The contrast enhancement is obtained by emphasizing the polarization rotation against the polarization properties of the isotropic background but the specimen must possess anisotropy. In Hoffman-Gross modulation contrast enhancement the modulator, a spatial intensity filter, is placed within the Fourier plane conjugate with a slit aperture. The image plane emphasizes phase gradients within the specimens and produces intensity variations proportional to the first derivative of the optical density within the object.\cite{HoffmanAO1975} 

In this work, we present a form of microscopy where the specimen is placed within an anisotropic object plane (anisotropic filter), thereby introducing the anisotropy contrast. We refer to this method as Anisotropic Contrast Optical Microscopy (ACOM). The specimen can be isotropic and/or anisotropic. The anisotropic filter can be composed of a transparent or highly reflective anisotropic thin film that is placed within the object plane. The contrast enhancement occurs within the images of the Mueller matrix\cite{MuellerMIT1943,JonesJOSA1947,Mueller1948} elements of the object plane. The Mueller matrix element images are obtained by principles similar to conventional Mueller matrix imaging instrumentation.\cite{footnote1RSIACOM2016} However, to increase accuracy the often exploited Fourier analysis algorithms\cite{Fujiwara_2007} are dispensed with in our ACOM instrumentation, and the Mueller matrix elements are obtained from a direct analysis of detected intensity data. Because the intrinsic contrast information in ACOM originates from a variation of the anisotropy in the object plane, our method differs from conventional Mueller matrix imaging\cite{Laude-BoulesteixSPIE2004,NovikovaOPNews2012,AntonelliOE2011,PierangeloOE2011,Laude-BoulesteixAO2004,FreudenthalChirality2009,LaraAO2006,JellisonApOpt_2006,ArteagaAO53_2014,ChenMMIRSI2016} or polarimetry imaging\cite{AzzamMMreview2016} methods.

In ACOM, the optical properties of the sample support -- the anisotropic filter -- cannot be ignored, in contrast to microscopy techniques which use isotropic supports. Before conclusions about certain optical and structural properties of a given specimen can be drawn, a good understanding of the optical properties of the anisotropic filter must be gained. Good understanding must also be gained about how these images change in the presence of a specimen. For example, for a given anisotropic filter, it is crucial to measure the actual Mueller matrix images of the anisotropic filter with and without a certain specimen. Such understanding can be gained from model calculations, on one hand, but also from experimental observation on the other hand. As we will show in this work, first, in ACOM the sensitivity to the presence of very small specimen is greatly enhanced compared with conventional microscopy methods. Thus, one may use ACOM for merely detecting presence of small (e.g., low-mass volume, highly transparent) specimen. Second, because of the ellipsometric model principles, attempts can be made to quantify the optical and structural properties of a specimen. The two arguments are the major advances of the ACOM over existing microscopy techniques. The concept of the instrumentation discussed here, therefore, bears potential for new imaging modalities in biomedical applications.

This paper is structured as follows. In Sec.~\ref{sec:Principle} we detail the principles of the ACOM concept, and the principles for calibration, operation, and quantitative image analyses for our ACOM instrumentation. In Sec.~\ref{sec:experimentalsetup} we describe a wavelength-tunable ACOM instrumentation. In order to illustrate the operation and functionality of our instrumentation, in Sec.~\ref{sec:ResultsandDiscussion} we demonstrate measurements on calibrated anisotropic filters with spatial patterns on isotropic transparent substrates. We then show and discuss images obtained after deposition of calibrated amounts of few-nm-sized particles into the void spaces of the anisotropic filter. We demonstrate the application in imaging geometries where the object plane is located within a microfluidic channel, we demonstrate use in imaging chromatography, and we monitor the adsorption of few-nm-thick organic layers. Finally, we present and discuss images obtained from mouse fibroblast cells, which are cultured onto the anisotropic filter.

\section{Principle}
\label{sec:Principle}

\begin{figure}[htb] \includegraphics[keepaspectratio=true,width=0.5\linewidth]{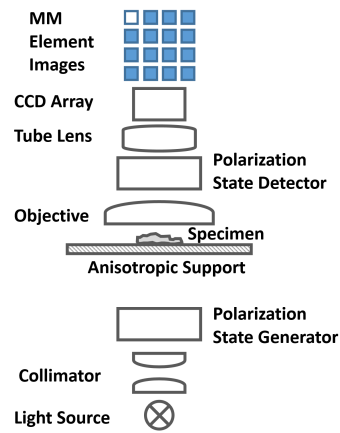}
\caption{Principle of an anisotropic contrast optical microscope in transmission configuration. The critical element is the anisotropic filter (support of specimen) placed within the object plane. Quantitative image analysis allows for deduction of up to 16 Mueller matrix element images. Ellipsometric model calculations of the Mueller matrix element images permit quantitative analysis of specimen properties, which is discussed in this work.}\label{fig:ACM}
\end{figure}

\subsection{ACOM}

The sample is illuminated under normal incidence as in traditional compound microscopes but the image forming light passages are polarization modulated (Fig.~\ref{fig:ACM}).\cite{footnote3RSIACOM2016} ACOM exploits ellipsometric operation principles. The principles are exploited for numerical inversion of the recorded intensity modulations for each individual pixel corresponding to a certain object area into Mueller matrix data, or into a specific ellipsometric sample model parameter.\cite{footnote2RSIACOM2016} An ellipsometric image processing approach allows to extract a set of non-redundant images, Mueller matrix images, from sets of multiple intensity images obtained under different polarization illumination and polarization detection conditions. The images contain up to 16 generally independent images of the individual elements of the Mueller matrix. The anisotropy variation caused by the presence of the sample within the object plane causes image contrasts in all 16 elements. The images reveal isotropic as well as anisotropic sample properties such as density and thickness, and birefringence and dichroism.

\subsection{Anisotropic Filter}

The illuminating light is split into two eigen polarization modes by the anisotropic filter, which interact upon transmission or reflection with the specimen. The eigen modes superimpose coherently afterward if the optical path lengths through specimen and filter are small against the coherence length. Depending on thickness, density, refractive index, and extinction coefficient of the specimen, as well as the specimen's anisotropy, the recombined light represents a slightly altered state of polarization relative to light that passes the anisotropic filter only, or light that passes the anisotropic filter and a different part of the specimen with different properties. The differences in polarization state cause the anisotropy contrast within the Mueller matrix images.\cite{footnote4RSIACOM2016} If the optical properties of the anisotropic filter are well known, attempts can then be made to analyze the anisotropy contrast within the Mueller matrix images, by searching quantitatively for optical properties of the specimen using ellipsometric model analysis procedures.

\begin{figure}[tbp]
\includegraphics[keepaspectratio=true,width=4cm, clip, trim=0 0 0 0 ]{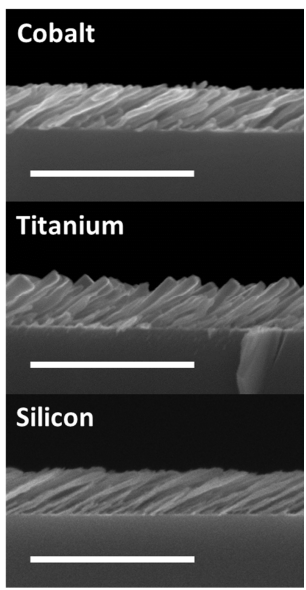}
\caption{Cross-section high-resolution scanning electron micrograph (SEM) images of slanted columnar thin films (SCTFs) from cobalt, titanium, and silicon, on crystalline silicon substrates. Scale bars are 500~nm. Similar SCTFs deposited onto glass slides are used in this work to establish the anisotropic filter in the ACOM instrumentation. See also Fig.~\ref{fig:STFpattern}. Image reprinted from Ref.~\onlinecite{SchmidtJAP114_2013} Reprinted with permission from American Institute of Physics.}
\label{fig:SCTFSEM}
\end{figure}

\subsection{SCTF}

\begin{figure}[tbp]
\includegraphics[keepaspectratio=true,width=7.5cm, clip, trim=0 0 0 0 ]{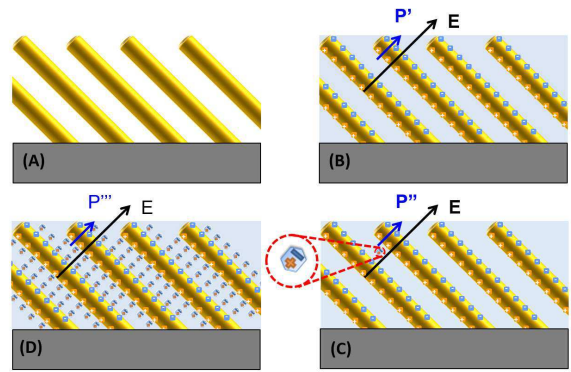}
\caption{Schematic presentation of shape-induced anisotropy in SCTF (A), which serve as anisotropic filters in the ACOM instrumentation. Changes of the shape-induced anisotropy caused by partial screening of the electric field, for example, due to immersion of the SCTF in a dielectric medium (B), and/or upon subsequent attachment of small molecules (C), or formation of coherently shaped overlayers (D). In each scenario, the dielectric polarizability for electric field polarization perpendicular to the columns is modified while the dielectric polarizability parallel to the columns remains nearly unaffected. In these scenarios, the amount of anisotropy is changed within the SCTFs.}
\label{fig:SCTFanisotropy}
\end{figure}

In this work, we have selected an anisotropic filter, which consists of nanoscopic, regularly shaped structures, which are arranged into a columnar thin film with highly coherently ordered nanostructures, and with a collective, slanted columnar direction. Such films, known as slanted columnar thin films (SCTF) possess strong anisotropic properties, which include birefringence and dichroism. Typical examples are shown in Fig.~\ref{fig:SCTFSEM}. SCTFs can be deposited by glancing angle deposition from a large variety of elemental and compound materials, such as Ti, Si, Al, Co, Ti$_2$O, Si$_2$O, Al$_2$O$_3$, etc.\cite{KundtAP1886,KoenigOptik1950,LakhtakiaMessierBook,HawkeyJVST2007,YoungNature1959,GoldsteinJAP1959,NieuwenhuizenPTR1966,RobbieJVST1995,RobbieJVST1995a,SchubertASSP46_2007,SchmidtPhD2010,BarrancoPMS2016} A subsequent deposition of a conformal, ultra-thin layer of metal-oxides, for example, by atomic layer deposition can modify the surface chemical properties of the SCTFs, and render them chemically inert under normal ambient.\cite{SchmidtAPL100_2012} The SCTFs used in this work possesses shape-induced birefringence,\cite{Hodgkinson_1998,Hodgkinson99} which is schematically shown in Fig.~\ref{fig:SCTFanisotropy}(A). The anisotropy is also crucially dependent on the fraction of the intercolumnar spacing as well as the diameter of the columns. The thickness of the SCTFs can be kept small against the wavelength at which the ACOM instrumentation may operate. Typically, the SCTFs are mostly empty, with $\approx$~75\% of void fraction.\cite{Schmidt_Bookchapter2013} 

\subsection{AB-EMA}
\label{sec:ABEMA}

For a SCTF, which renders a highly ordered topography of anisotropic inclusions the Bruggeman effective medium approximation\cite{BruggemanAP24_1935} can be modified by introducing depolarization factors $L^{\mathrm{D}}_{n,j}$ ($j=a,b,c$) along each of the three major SCTF optical polarizability axes (\textbf{a}, \textbf{b}, and \textbf{c}) for the $n$th component (See, for example, Refs.~\onlinecite{HofmannAPL99_2011,SchmidtPhD2010,Schmidt_Bookchapter2013,RodenhausenChapter2014SCTF}). The depolarization factors render the now anisotropic polarizability-describing inclusions as ellipsoidal.\cite{Polder1946} Thus, three effective dielectric function components, each averaged over the respective polarizability axis, are determined. The anisotropic Bruggeman effective medium approximation (AB-EMA) equations for $m$ constituent materials are:

\begin{eqnarray}
\label{Bruggeman}
\sum_{n=1}^m f_n = 1, \\
\sum_{n=1}^m f_n\frac{\varepsilon_n-\varepsilon_{\mathrm{eff},j}}
{\varepsilon_{\mathrm{eff},j}+L^{\mathrm{D}}_{n,j}\left(\varepsilon_n-\varepsilon_{\mathrm{eff},j}\right)}=0, \; j=a,b,c,
\end{eqnarray}

\noindent where $\varepsilon_{\mathrm{eff},j}$ is the effective dielectric function along the $j$th axis, $\varepsilon_n$ is the bulk dielectric function of the $n$th constituent material, and $f_n$ is the volume fraction of the $n^{th}$ material.\cite{Granqvist1996,HofmannAPL99_2011,SchmidtPhD2010,Schmidt_Bookchapter2013,RodenhausenChapter2014SCTF}

\subsection{Screening}

Figures~\ref{fig:SCTFanisotropy}(B)-(D) schematically depict scenarios of different modifications of the anisotropy of an SCTF, for example during immersion within a liquid, or after attachment of organic adsorbates. The modifications are due to partial screening of the anisotropic polarization charges within the columnar structures. In general, attachment of dielectric or conductive materials onto the colum surfaces and/or inclusion of material into the intercolumnar void space changes the anisotropic optical constants. We have previously demonstrated that SCTFs can be used to sensitively detect attachment or desorption of very small amounts of organic substances within the empty space of SCTFs.\cite{RodenhausenOE20_2012,KasputisJCC_2013,RodenhausenChapter2014SCTF,KoenigABC2014} The extreme sensitivity to such small attachments originates from the fact that generalized ellipsometry is extremely sensitive to small changes in cross-polarization properties of thin films and surfaces.\cite{footnote5RSIACOM2016}

\subsection{Ellipsometry}

Principles of Mueller matrix ellipsometry are used in this work for image generation from series of measured intensity images. The Stokes vector $\mathbf{S}$ description is used here within the traditional $p$-$s$ coordinate system.\cite{footnote6RSIACOM2016} For the ACOM instrumentation the interaction of light is considered for normal incidence only, and the plane of incidence is not defined. However, we assign a fixed direction within the instrumentation as the $p$ direction by choice of parameters. All images are lined up with the $p$-$s$-coordinate system, and the optical axes of the anisotropic filter are set relative to $p$. The real-valued $4\times 4$ Mueller matrix $\mathbf{M}$ describes the change of electromagnetic plane wave properties (intensity, polarization state),\cite{goldstein2003polarized,JarrendahlMuellerJANews2011} expressed by a Stokes vector $\mathbf{S}$, upon change of the coordinate system or the interaction with a sample, with an optical element, or any other matter :\cite{Fujiwara_2007,AzzamBook_1984}

\begin{equation}
	S^{(\text{out})}_j=\sum^{4}_{i=1}M_{ij}S^{(\text{in})}_i,\;\;(j=1\ldots 4)\;,
	\label{eqn:multi_MM}
\end{equation}

\noindent where $\mathbf{S}^{(\text{out})}$ and $\mathbf{S}^{(\text{in})}$ denote the Stokes vectors of the electromagnetic plane wave before and after the change of the coordinate system, or an interaction with a sample, respectively. Note that all Mueller matrix elements presented in this work are normalized by the element $M_{11}$, therefore $|M_{ij}|\leq 1$ and $M_{11}=1$. Experimental determination of the Mueller matrix, or selected elements of the Mueller matrix is often termed generalized (Mueller matrix) ellipsometry,\cite{AzzamBook_1984,SchubertJOSAA13_1996,Schubert03h,Schubert04,JellisonHOE_2004,Fujiwara_2007,JellisonPRB2011CdWO4,SchubertPRB2016} or Mueller matrix polarimetry.\cite{Goldstein1990,Goldstein1992,goldstein2003polarized} Numerous approaches exist.\cite{Hauge:78,AzzamOL1978MMpol,GilMSThesis1979,HaugeSS96_1980,AzzamOA1982,BernabeuJO1985,Goldstein1992,Jellison98a,Jellison98b,LeeRSI_2001,Chen03,DeMartinoOL2003LCMM,Laude-BoulesteixAO2004,Laude-BoulesteixSPIE2004,LaraAO2006,TuchinBook2006,IchimotoPNAO2006,Fujiwara_2007,BenHatitpssa2008,FreudenthalChirality2009,ArsTSF2011,AntonelliOE2011,GhoshJBO2011,PierangeloOE2011,ArteagaAO2012,NovikovaOPNews2012,ArteagaAO53_2014,AzzamMMreview2016} Important characteristics of a given instrumental approach is the incorporation of two sets of light polarization and polarization mode phase shifting components. Such can be selected from sets of linear and circular polarizers, for example. Depending on whether these elements are included or not, certain rows and/or certain columns of the Mueller matrix may not be accessible. In our instrumentation, the fourth column is not accessible due to the lack of a second compensator.\cite{Fujiwara_2007,footnote7RSIACOM2016}

\subsection{Image analysis}
\label{sec:imageanalysis}

\begin{figure}[tbp]
\includegraphics[keepaspectratio=true,width=7.1cm, clip, trim=0 0 0 0 ]{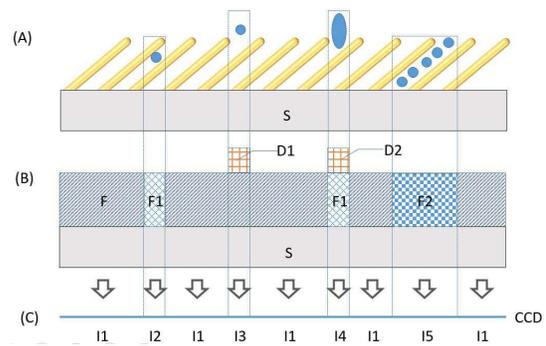}
\caption{Schematic presentation of best-match-model parameter image calculation using predefined sets of most likely model scenarios. In (A), different scenarios of adsorbate attachment within the open space of SCTFs are shown. In (B), these scenarios are translated into multiple layer model calculations, for example, determining the adsorbate fraction (F, F1, F2, ...), or thickness of an additional over layer (D1, D2, ...). In (C), image formation in the ACOM instrumentation is shown schematically at the detector, where certain areas may contain pixels with similar or equal information (I1, I2, ...). Such may be collated into one effective pixel (``binning''), if needed. For each effective pixel, model calculations may reveal sets of most likely parameters, for example, fraction F and thickness D.}
\label{fig:OACOMimagemodel}
\end{figure}

In ACOM, different types of images are determined: (i) images of polarized intensities, (ii) images of Mueller matrix elements, and (iii) images of ellipsometric model parameters. Images (ii) are obtained through model calculations from images (i). Images (iii) are either obtained by model calculations targeting best-match to images (i), or by different model calculations targeting best-match to images (ii). Such model parameters are, for example, the thickness of a thin layer, or the optical constants of a constituent of the specimen, or structural parameters such as azimuthal orientation of optical axes, etc.  The specimen under investigation can be modeled using a multiple layer approach, where a stack of homogeneous layers with assumed ideal, plane-parallel interfaces is located on a substrate. Here the substrate material is considered transparent, and modeled with the refractive index of glass (BK7).\cite{Palik98a} On top of the substrate the anisotropic filter is modeled as an anisotropic thin film with thickness $d_{\mathrm{SCTF}}$. The thickness of this layer $d_{\mathrm{SCTF}}$ is chosen here smaller than the wavelength, and may range from few nm to few hundred nm. A second layer may be considered for dielectric or absorbing specimens, which are supported by the anisotropic filter. This second layer may be anisotropic as well, if the specimen is anisotropic. A 4$\times$4 matrix formalism is then used.\cite{Schubert96,Schubert03h,Schubert04} Data analysis requires nonlinear regression methods, where measured and calculated data are matched as close as possible by varying appropriate physical model parameters.\cite{JellisonHOE_2004} Thorough discussions of proper data modeling in ellipsometry can be found in the literature, for example, reviews are provided in Refs.~\onlinecite{Tompkins_2004,HumlicekHOE,SchubertIRSEBook_2004,SchubertADP15_2006,Fujiwara_2007}

In Fig.~\ref{fig:OACOMmagemodel}A, scenarios are shown when individual molecules adsorb within the open space of the SCTF, or partially within and outside, or completely fill the void fraction. The best-match-model comprises a set of possible model layer situations (Fig.~\ref{fig:OACOMmagemodel}B) for every pixel. Certain pixels of the detector area may be collated into one effective pixel (``binning''). For every effective pixel, a certain number of possible model layer calculations may be performed, and the best match model parameter(s) may be decided as the most likely physical circumstance of the image forming anisotropic filter properties. For example, images can be obtained which contain the surface volume-mass density of an organic adsorbate over an area within the object plane. 

\section{Experimental Setup}
\label{sec:experimentalsetup}

\subsection{Instrumentation description}
\label{sec:instrumentationdescription}

\begin{figure*}[tbp]
\includegraphics[keepaspectratio=true,width=0.8\textwidth, clip, trim=0 0 0 0 ]{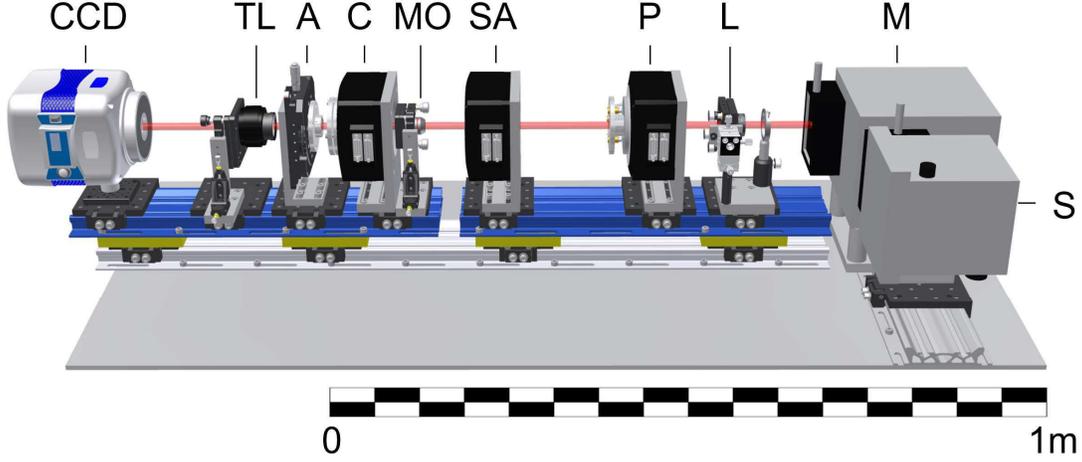}
\caption{Technical drawing (to scale) of the ACOM instrumentation. Also indicated is the optical beam path. The instrument is equipped with a short arc lamp as light source (S) and grating monochromator (M), which permits operation at tunable wavelength in the range from 300~nm to 1000~nm. The system is further composed of a collimation lens (L), a polarizer (P), a sample stage (SA), an infinity-corrected-microscope objective (MO), a compensator (C), an analyzer (A), a tube lens (TL), and an imaging detector (CCD). Parts CCD, TL, A, C, and MO are mounted onto a common rail to allow for convenient image position correction when MO is replaced for variation of focal length in order to obtain different lateral magnification.}\label{fig:Instrument}
\end{figure*}

The ACOM instrumentation presented here operates in a polarizer--sample--compensator--analyzer configuration. The polarizer, the compensator, and the analyzer can be rotated by azimuthal increments. Hence, the instrumentation provides images of Mueller matrix elements except for elements in the 4$^{th}$ column.\cite{Fujiwara_2007} The instrument permits tunable wavelength ellipsometric measurements in the wavelength range from 300~nm to 1000~nm. In some cases, inclusion of data at more than one wavelength provides additional information of the specimen under investigation. Images when analyzed at multiple wavelengths can improve uncertainty limits on model parameters obtained from data analysis. Variation of wavelength could also be used to increase sensitivity to certain model parameters, and which is not further discussed in this work. For example, the sensitivity of the ACOM instrumentation to the presence of small organic adsorbates in the open volume of SCTFs strongly depends on wavelength, and which will be the subject of forthcoming work. The experimental setup of the ACOM discussed here is based on a normal incidence transmission arrangement. Imaging of the specimen within the object plane is performed using objective and tube lens arrangements as discussed further below. A drawing of the ACOM instrumentation is shown in Fig.~\ref{fig:Instrument}.

A 100 W mercury arc lamp is employed as the light source (S). The light is passed through a dual-grating imaging monochromator (M; Princeton Instruments sp-150). The latter is equipped with two gratings blazed for 500 nm, with 300 and 600 lines/mm, respectively. Part S is directly mounted onto the monochromator entrance slit. The monochromatic light emitted from the exit slit of the monochromator is then collimated by a 100-mm-focal-length, 1'' diameter, achromatic-doublet-collimation lens (L). A Glan-Thompson polarizer (P) is used to control the incident polarization state. Rotation of the polarizer P by azimuth angle $\theta_{\mathrm{P}}$ is achieved by a high-precision, motorized-rotation stage (Newport RGV100BL). The same type of rotation stage is used to support the sample stage (SA), which allows automated execution of ACOM data acquisition as a function of sample rotation azimuth. The sample stage supports the anisotropic filter. The anisotropic filter comprises a transparent microscope slide (BK7) with a SCTF deposited onto one side, as shown in Fig.~\ref{fig:SCTFanisotropy}(A). The light after interacting with the anisotropic filter/sample is then collected by an infinity-corrected-microscope objective (MO). The light is then passed through a compensator (C), mounted onto the same type of motorized-rotation stage as for parts P and SA (azimuth parameter $\theta_{\mathrm{C}}$). Another Glan-Thompson polarizer is used as the analyzer (A). The analyzer is mounted into a manual-rotation stage (azimuth parameter $\theta_{\mathrm{A}}$). Individual rays of light that leave the object plane form an image on the detector (D) through an apochromatic tube lens (TL; Thorlabs ITL200). The working distance of the tube lens is 148~mm. Detector D is established by a low-noise, charge-coupled-device (CCD) camera (Photometrics Evolve 512 Delta). The magnification of lateral distances between objects in the object plane (anisotropic filter) is a function of the magnification of the objective lens MO, and which can be adjusted for a given experimental requirement by replacing MO with a different magnification. Standard infinity-corrected objective lenses can be used. During the adjustment, parts CCD, TL, A, C, and MO are moved together along the optical axis to accommodate for the correct image position of MO relative to SA. For this purpose, parts CCD, TL, A, C, and MO are mounted onto a common rail. This rail is mounted onto a base rail onto which all remaining parts are mounted (Fig.~\ref{fig:Instrument}).

\subsection{Anisotropic filter}
\label{sec:anisotropicfilter}

The anisotropic filter in the ACOM instrumentation consists of a semi-transparent SCTF, which is deposited by GLAD onto transparent microscope slides (glass substrates).  An in-house built GLAD deposition system is used.\cite{SchmidtPhD2010,SchubertASSP46_2007} The glass substrates are purchased from Lakeside Microscope Accessories. The thickness parameters and the slanting angle of the SCTFs can be controlled by growth conditions. Details of specific SCTFs used in this work are described in the application sections further below. The SCTFs possess strong optical anisotropy, including a strong, wavelength dependent dichroism and birefringence. The SCTFs are optically anisotropic, and must be described by three effective major optical constants.\cite{SchmidtAPL94_2009,SchmidtOL34_2009} Generalized spectroscopic ellipsometry (GSE)\cite{Azzam72,SchubertJOSAA13_1996,Jellison98a} is demonstrated as a suitable approach to accurately characterize the anisotropic optical properties of SCTFs.\cite{SchmidtAPL94_2009,SchmidtOL34_2009} A series of recent publications have reported on GSE investigations for a variety of SCTFs prepared from dielectric, semiconductor, metallic, and magnetic materials.~\cite{SchmidtJAP105_2009,SchmidtAPL96_2010,SchmidtTSF519_2011,SchmidtAPL100_2012,LiangAPL2013SCTFpolymer,Schmidt_Bookchapter2013,SchmidtJAP114_2013,WilsonJMCC2013,WilsonAPL2015graphene}

\begin{figure*} \includegraphics[keepaspectratio=true,width=\textwidth, clip, trim=0 0 0 0 ]{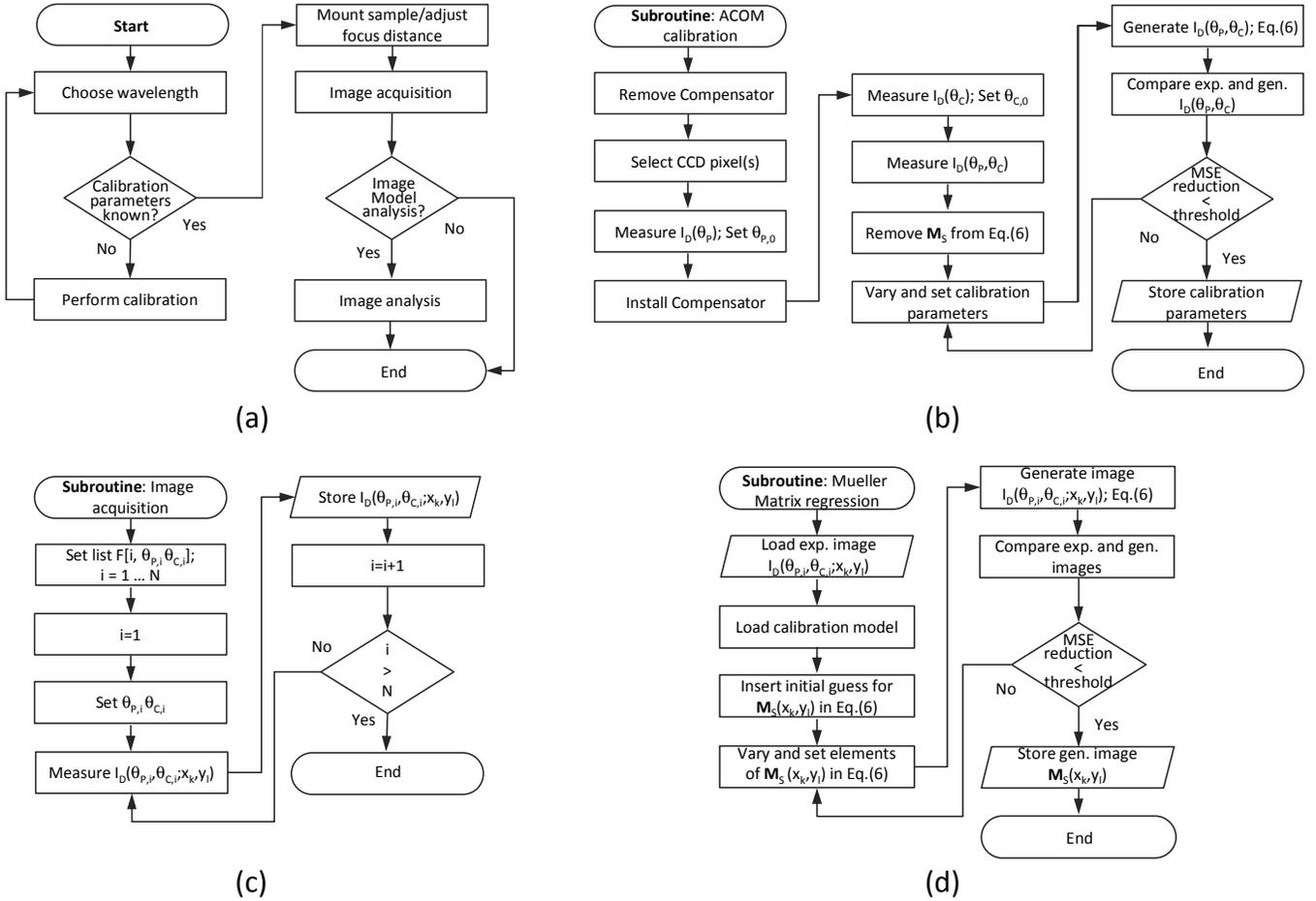}
\caption{Flow chart of the ACOM instrumentation data acquisition and calibration process. (a) Shows the basic ACOM operation. (b) Summarizes the ACOM instrumentation calibration procedure. (c) Details the principle operation for ACOM image acquisition. (d) Depicts the Mueller matrix element regression process for transforming ACOM images into Mueller matrix images.}\label{fig:ACOMflowchart}
\end{figure*}

\subsection{Instrumentation calibration and operation}
\label{sec:instrumentationcalibrationandoperation}

Figure~\ref{fig:ACOMflowchart}(a) depicts a flow chart for operation of the ACOM instrumentation. Initially, a wavelength is selected. Prior to performing measurements a calibration process is followed (see Sec.~\ref{sec:calibration}), Fig.~\ref{fig:ACOMflowchart}(b). Once the instrumentation is calibrated, a sample/specimen is mounted onto the anisotropic filter. This process may also involve a liquid or gaseous flow cell, for example in a microfluidic device encapsulated between transparent glass slides (see sec.~\ref{sec:ACOMCTAB}). Then image acquisition follows (Fig.~\ref{fig:ACOMflowchart}(c)). This process is explained in Sec.~\ref{sec:acquisition}. The images can then be stored as the final result of the operation, or the images can be further analyzed by model calculations. Figure~\ref{fig:ACOMflowchart}(d) depicts one example, where the images are analyzed by a best-match-model calculation to obtain the images of the Mueller matrix elements (Sec.~\ref{sec:MMregression}). At the beginning, all optical components are physically aligned along the optical axis as accurately as possible. Precision engineered common rails and element supports ensure high mounting accuracy and stability.

\subsubsection{Calibration}
\label{sec:calibration}

The goal of the calibration is to obtain best-match-model parameters (calibration constants), which describe the polarization properties of all polarizing elements within the ACOM instrumentation. These parameters are required during operation and image analysis of the instrumentation. As the first step, the analyzer is set to a fixed position. The polarization direction of the analyzer thereby sets the $s$ direction, subsequently defines the $p$ direction, and hence inscribes the ACOM coordinate system within which the obtained Mueller matrix images will be cast. 

\paragraph{Calibration model:} A chain model of Mueller matrices can be used to describe the detected intensity for each pixel. The Stokes vector at the detector (CCD) in Fig.~\ref{fig:Instrument} is obtained by the ordered product of the Mueller matrices of polarizer $\mathbf{M}_{\mathrm{P}}$, sample $\mathbf{M}_{\mathrm{S}}$, compensator $\mathbf{M}_{\mathrm{C}}$, and analyzer $\mathbf{M}_{\mathrm{A}}$, and matrices $\mathbf{R}$ to account for azimuth rotations\cite{Tompkins_2004,Fujiwara_2007}
 
\begin{widetext}
\begin{equation}
I_{\mathrm{D0}}\mathbf{s}_{\mathrm{D}}=\mathbf{M}_{\mathrm{A}}\mathbf{R}(-\theta_{\mathrm{C}})\mathbf{M}_{\mathrm{C}}(\delta)\mathbf{R}(\theta_{\mathrm{C}})\mathbf{M}_{\mathrm{S}}\mathbf{R}(-\theta_{\mathrm{P}})\mathbf{M}_{\mathrm{P}}\mathbf{R}(\theta_{\mathrm{P}})I_{\mathrm{S}}\mathbf{s}_{\mathrm{S}}, 
\label{eq:MMchainACMI}
\end{equation}
\end{widetext}
 
\noindent where the normalized Stokes vectors and the irradiance at the source (detector) are denoted by $I_{\mathrm{S}}$ ($I_{\mathrm{D0}}$) and $\mathbf{s}_{\mathrm{S}}$ ($\mathbf{s}_{\mathrm{D}}$), respectively, and:

\begin{equation}
\mathbf{\bm{M}_i}=\frac{1}{2}
\begin{bmatrix}
		\hspace{8pt}X_{i,11}&\hspace{17pt}X_{i,12}&\hspace{17pt}0&\hspace{18pt}0&\hspace{5pt}{}\\
		\hspace{8pt}X_{i,12}&\hspace{17pt}X_{i,22}&\hspace{17pt}0&\hspace{18pt}0&\hspace{5pt}{}\\
		\hspace{8pt}0&\hspace{17pt}0&\hspace{17pt}0&\hspace{18pt}0&\hspace{5pt}{}\\
		\hspace{8pt}0&\hspace{17pt}0&\hspace{17pt}0&\hspace{18pt}0&\hspace{5pt}{}
\end{bmatrix},
\label{eq:MMPol}
\end{equation}

\begin{equation}
\mathbf{\bm{M}_C}=
\begin{bmatrix}
		\hspace{8pt}1&\hspace{17pt}0&\hspace{10pt}0&\hspace{0pt}0\\
		\hspace{8pt}0&\hspace{17pt}1&\hspace{10pt}0&\hspace{0pt}0\\
		\hspace{8pt}0&\hspace{17pt}0&\hspace{8pt}\cos \delta&\hspace{2pt}\scalebox{0.6}[1.0]{\(-\)}\sin \delta\\
		\hspace{8pt}0&\hspace{17pt}0&\hspace{10pt}\sin \delta&\hspace{7pt}\cos \delta
\end{bmatrix},
\label{eq:MMComp}
\end{equation}

\begin{equation}
\mathbf{R}\left(\theta_j\right)=
\begin{bmatrix}
		\hspace{8pt}1&\hspace{6pt}0							&\hspace{-12pt}0																&\hspace{1pt}0&\hspace{5pt}{} \\
		\hspace{8pt}0&\hspace{4pt}\cos 2\theta_j	&\hspace{-1pt}\sin 2\theta_j									&\hspace{1pt}0&\hspace{5pt}{} \\
		\hspace{8pt}0&\hspace{-2pt}\scalebox{0.6}[1.0]{\(-\)}\hspace{1pt}\sin 2\theta_j				&\hspace{-1pt}\cos 2\theta_j						&\hspace{1pt}0&\hspace{5pt}{} \\
		\hspace{8pt}0&\hspace{7pt}0							&\hspace{-12pt}0																&\hspace{1pt}1&\hspace{5pt}{} 
\end{bmatrix},
\label{eq:MMRotation}
\end{equation}

\begin{equation}
\mathbf{M_S}=
\begin{bmatrix}
		\hspace{3pt}M_{11}&\hspace{4pt}M_{12}&\hspace{4pt}M_{13}&\hspace{4pt}M_{14}&\hspace{-2pt}{}  \\
		\hspace{3pt}M_{21}&\hspace{4pt}M_{22}&\hspace{4pt}M_{23}&\hspace{4pt}M_{24}&\hspace{-2pt}{}  \\
		\hspace{3pt}M_{31}&\hspace{4pt}M_{22}&\hspace{4pt}M_{33}&\hspace{4pt}M_{34}&\hspace{-2pt}{}  \\
		\hspace{3pt}M_{41}&\hspace{4pt}M_{42}&\hspace{4pt}M_{43}&\hspace{4pt}M_{44}&\hspace{-2pt}{} 
\end{bmatrix},
\label{eq:MMSample}
\end{equation}

\noindent where $i$=``$\mathrm{P}$'', ``$\mathrm{A}$''. The parameters $\theta_j$ ($j$=``$\mathrm{C}$'', ``$\mathrm{P}$'') describe the azimuth orientation of compensator and polarizer rotations.\cite{HofmannRSI77_2006} The parameter $\delta$ is the relative phase shift (retardation) between fast and slow axes of the compensator C, and which may depend on wavelength. Parameters $X_{i,11}, X_{i,12}, X_{i,22}$ account for nonideality of P and A. For an ideal polarizing element, $X_{i,11}=X_{i,12}=X_{i,21}= X_{i,22}=1$. For a nonideal polarizer, these parameters can be less than unity. 

The polarization properties of the normalized Stokes vector at the exit slit of the monochromator (source) are described by chain multiplication of an ideal polarizer and compensator Mueller matrix, and the Stokes vector for unpolarized light:\cite{goldstein2003polarized}

\begin{equation}
I_{\mathrm{S}}\mathbf{s}_{\mathrm{S}}=\mathbf{M}_{\mathrm{C,s}}(\delta_{\mathrm{s}})\mathbf{R}(-\theta_{\mathrm{s}})\mathbf{M}_{\mathrm{P,s}}\mathbf{R}(\theta_{\mathrm{s}})\left(1,0,0,0\right)^{T}, 
\label{eq:sourcepol}
\end{equation}

\noindent where $\theta_{\mathrm{s}}$ denotes the source polarization azimuth, $\delta_{\mathrm{s}}$ is the source polarization phase shift, and $T$ denotes the transpose of a vector.

The signal detected at the CCD, $I_{\mathrm{D0}}$, may be affected by a nonlinear detector response. Ideally, a detector responds to a linear increase in irradiance (power/area) with a linear increase in electrical signal. The response of a nonideal detector, $I_{\mathrm{D}}$, is described here as: 

\begin{equation}
  I_{\mathrm{D}}=(\alpha+\beta I_{\mathrm{D0}}+\gamma I^2_{\mathrm{D0}})I_{\mathrm{D0}},
\label{eq:nonlinearcorr}
\end{equation}

\noindent where $\alpha$ is the linear response coefficient, and $\beta$ and $\gamma$ are the first and second-order nonlinearity correction coefficients of the CCD detector. Hence, $I_{\mathrm{D0}}$ in Eq.~(\ref{eq:MMchainACMI}) is replaced with $I_{\mathrm{D}}$. Note that we treat the response of the detector insensitive to polarization. Parameters to be determined during the calibration process (calibration parameters) are for polarizer and compensator: $\theta_{j,0}$,  $X_{i,11}, X_{i,12}, X_{i,22}$, $\delta$, for detector: $\alpha, \beta, \gamma$, and for source: $\theta_{\mathrm{s}}$, $\delta_{\mathrm{s}}$.

\paragraph{Calibration process:} The signal is obtained by measuring and recording intensity through the instrument with the sample/specimen removed, either for all pixels individually or averaged over certain pixel areas. We use 140 $\times$ 140 pixels within the center area of the available 512 $\times$ 512 pixels of the CCD camera. We assume that all optical elements are homogeneous across the relevant beam diameter. This is ensured by selecting optical elements whose effective aperture is sufficiently large compared to the effective beam diameter ($\approx$ 10~mm). The relevant, or effective beam diameter circumscribes the area across which light entering the pixel area of the detector is traversing all optical elements. It is thereby also assumed that the detector response function is equal for all pixels. The procedure must be repeated for every wavelength at which the instrumentation is used to obtain Mueller matrix images, and parameters are determined as a function of wavelength. In an improvement step, a refined procedure may repeat the calibration described here by allowing the process in Fig.~\ref{fig:ACOMflowchart}(b) to be evaluated for every pixel individually, and by determining all calibration parameters as a function of pixel index. Such a procedure is not performed in this work.

Initially, element C is removed. In this first step, an initial, best estimate for the angular azimuth parameters of the polarization directions for P relative to A as well as for the fast and slow axis orientations of C is obtained. Then, the azimuth angle parameters are $\theta_i=\theta_{i,0}+\theta_{i,m}$ ($i$=``$\mathrm{P}$'', ``$\mathrm{C}$''), where $\theta_{i,m}$ is the angular increment progressed by the motorized stages, and $\theta_{i,0}$ is the offset angle. P is rotated in increments of 2$^{\circ}$ from $0^\circ \dots 180^\circ$, and the signal is recorded versus $\theta_P$. A simple minimum-search procedure then allows to identify, as an initial, best estimate, the offset angle $\theta_{\mathrm{P,0}}$. Next, C is inserted into the system. The polarizer is then positioned at $\theta_{\mathrm{P}}=90^\circ$, that is, nearly crossed to A. Then, C is rotated step-wise in increments of 2$^{\circ}$ from $0^\circ \dots 180^\circ$ and the signal is recorded as a function of $\theta_{\mathrm{C}}$. When the fast axis of C is aligned parallel with the polarization direction of P, the state of polarization of light transmitted through P and C remains unaffected (``nulling position''). Hence, a simple minimum search procedure then allows to identify, as an initial, best estimate, $\theta_{\mathrm{C,0}}$. Then, the calibration continues with acquisition of intensity data as a function of $\theta_{\mathrm{P}}$ and $\theta_{\mathrm{C}}$, for $\theta_{\mathrm{P}} = 0 ^\circ \dots 180^\circ$ in 3$^{\circ}$ increments, and for $\theta_{\mathrm{C}} = 0 ^\circ \dots 180^\circ$ in 10$^{\circ}$ increments. The list of $I_{\mathrm{D}}(\theta_{\mathrm{P}},\theta_{\mathrm{C}})$ with 120$\times$36 data points (sets) is stored, and then analyzed by a best-match model parameter calculation. Experimental and calculated data $I_{\mathrm{D}}(\theta_{\mathrm{P}},\theta_{\mathrm{C}})$ are compared, and parameters $\theta_{j,0}$,  $X_{i,11}, X_{i,12}, X_{i,22}$, $\delta$, $\alpha, \beta, \gamma$, $\theta_{\mathrm{s}}$, and $\delta_{\mathrm{s}}$ are varied until best match is obtained. A weighed error sum is used, where systematic experimental uncertainties values are incorporated into the numerical regression algorithm. As a result, the best-match-model calibration parameters are obtained with numerically estimated uncertainty limits.

\subsubsection{Image acquisition}
\label{sec:acquisition}

A list of $i$ settings for polarizer ($\theta_{\mathrm{P}}$) and compensator azimuth ($\theta_{\mathrm{C}}$) positions, F[$i,\theta_{\mathrm{P},i},\theta_{\mathrm{C},i}$], is determined (Fig.~\ref{fig:ACOMflowchart}(c)). This list may contain large numbers of entries, $N$. A priori, no criterion exists which settings to include. In general, an experiment should cover as much as possible of the two-dimensional area in $\theta_{\mathrm{P}}$ and $\theta_{\mathrm{C}}$, and in sufficient detail. Hauge, and Jellison and Modine suggested a Fourier analysis, and minimum settings were discussed which must satisfy the  Nyquist criterion.\cite{Hauge:78,HaugeSS96_1980,Jellison98b} Model calculations predicting the shape of $I_{\mathrm{D}}(\theta_{\mathrm{P}},\theta_{\mathrm{C}})$ for a given anisotropic filter and specimen may help identifying best conditions. Such conditions are when $I_{\mathrm{D}}(\theta_{\mathrm{P}},\theta_{\mathrm{C}})$ reveals strongest changes with placement of the sample/specimen. See Sec.~\ref{sec:ACOMCTAB} for a discussion of suitable settings in a specific case. Acquisition of $N$ images is then performed by detecting and storing images of the CCD detector, which may be addressed by pixel arguments $1 \dots k \dots 512$ and $1 \dots l \dots 512$, and stored for each setting of $\theta_{\mathrm{P}}$ and $\theta_{\mathrm{C}}$ prescribed within the list F, respectively. For each image, the experimental uncertainties for each pixel is stored as well. The experimental uncertainties are determined as the systematic error of the pixel values delivered by the CCD camera. Typical acquisition time for one single image is 10~ms. Typical times for performing a set of images in a list such as F[$i,\theta_{\mathrm{P},i},\theta_{\mathrm{C},i}=3\theta_{\mathrm{P}}$], with increments in P by 3$^{\circ}$ over one full rotation, is 45~s.

\subsubsection{Mueller matrix regression}
\label{sec:MMregression}

After successful calibration, and after image acquisition, images $I_{\mathrm{D}}(\theta_{\mathrm{P}},\theta_{\mathrm{C}})$ can be analyzed using Eqs.~(\ref{eq:MMchainACMI})-(\ref{eq:nonlinearcorr}). All Mueller matrix elements of $\mathbf{M}_{\mathrm{S}}$ are considered as model parameters. For each pixel Eq.~(\ref{eq:MMchainACMI}), with all necessary calibration parameters, is used to calculate $I_{\mathrm{D}}(\theta_{\mathrm{P}},\theta_{\mathrm{C}})$ for all polarizer and compensator settings prescribed in list of sets F[$i,\theta_{\mathrm{P},i},\theta_{\mathrm{C},i}$] (Fig.~\ref{fig:ACOMflowchart}(d)). The calculated and experimental data are compared. A regression analysis procedure is used to minimize the mean square error function (MSE), which is weighed by the experimental uncertainties for each data point. In the regression procedure, only the Mueller matrix elements of the sample/specimen are varied. The result of the regression procedure is the set of images of Mueller matrix elements.

\subsubsection{Ellipsometric model parameter regression}
\label{sec:EMregression}

Images $I_{\mathrm{D}}(\theta_{\mathrm{P}},\theta_{\mathrm{C}})$ can be analyzed using Eqs.~(\ref{eq:MMchainACMI})-(\ref{eq:nonlinearcorr}) with all Mueller matrix elements in $\mathbf{M}_{\mathrm{S}}$ obtained by using an ellipsometric model calculation. In the ellipsometric model calculation, the Mueller matrix elements are calculated using multiple layered models. An example is discussed above in Sec.~\ref{sec:imageanalysis}. Hence, one can obtain images of model parameters, for example, thickness of layers, index of refraction of layers, etc. Mueller matrix element images can still be obtained, but these calculated images then contain the constraints of the physical model used for their calculation. The best-match-model parameter images can be very useful when basic parameters such as thickness or surface mass area density are of primary concern. The advantage lies in the substantial reduction in number of images for the sample of interest, for example, one surface area mass density image versus 11 Mueller matrix images. 


\section{Results and Discussion}
\label{sec:ResultsandDiscussion}

We present demonstrations of the ACOM instrumentation for its performance to measure Mueller matrix images of calibrated, patterend anisotropic filters. We discuss lateral resolution calibration on well-characterized features of anisotropic materials. We demonstrate detection of ultra-thin-layer formation within calibrated areas of anisotropic filters within a liquid cell. The process of organic and inorganic layer attachment onto SCTFs is well characterized in previous work.\cite{SchmidtTSF519_2011,RodenhausenOE20_2012,SchmidtAPL100_2012,KasputisJCC_2013,LiangAPL2013SCTFpolymer,KoenigABC2014,RodenhausenChapter2014SCTF,LiangAPLporousSCTF,WilsonAPL2015graphene,KasputisABM2015} From our previous results, we derive here a new sensitivity limit for laterally-resolved detection of surface mass density with our ACOM instrumentation. We then present a quantitative measurement of laterally-resolved surface mass density of titanium dioxide nanoparticles dispensed within predefined areas of a calibrated anisotropic filter. Deposition of the nanoparticles is made with a commercially-available, volume-calibrated nanoplotter instrumentation. Hence, we compare quantitative measurements obtained with the ACOM instrumentation and the exact density of nanoparticles dispersed onto the sample surface. We finally demonstrate the ACOM instrumentation for imaging of test dye chromatographic flow separation, and for imaging of living cells, which are cultured onto the anisotropic filter.  

\subsection{ACOM on patterned anisotropic filters}
\label{sec:Nimaging}

\begin{figure}[tbp] \includegraphics[keepaspectratio=true,width=8.2cm, clip, trim=0 0 0 0]{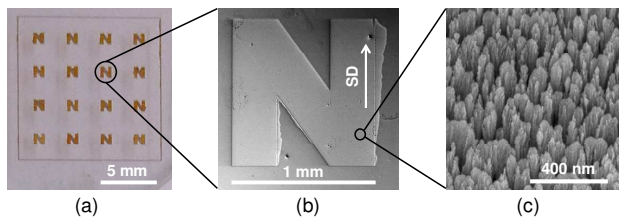}
\caption{(a) Photographic image of a patterned Si-SCTF sample deposited using silicon onto a transparent (BK7) microscope slide. SEM images shown in (b) abd (c) are taken over the surface area of the patterned Si-SCTF sample. The slanting (SD) direction is indicated in (b). The patterned areas act as anisotropic filters with calibrated lateral extensions, and serve for calibration of the lateral ACOM image scales.}
\label{fig:STFpattern}
\end{figure}

\begin{figure*}[tbp] \includegraphics[keepaspectratio=true,width=\textwidth, clip, trim=0 0 0 0 ]{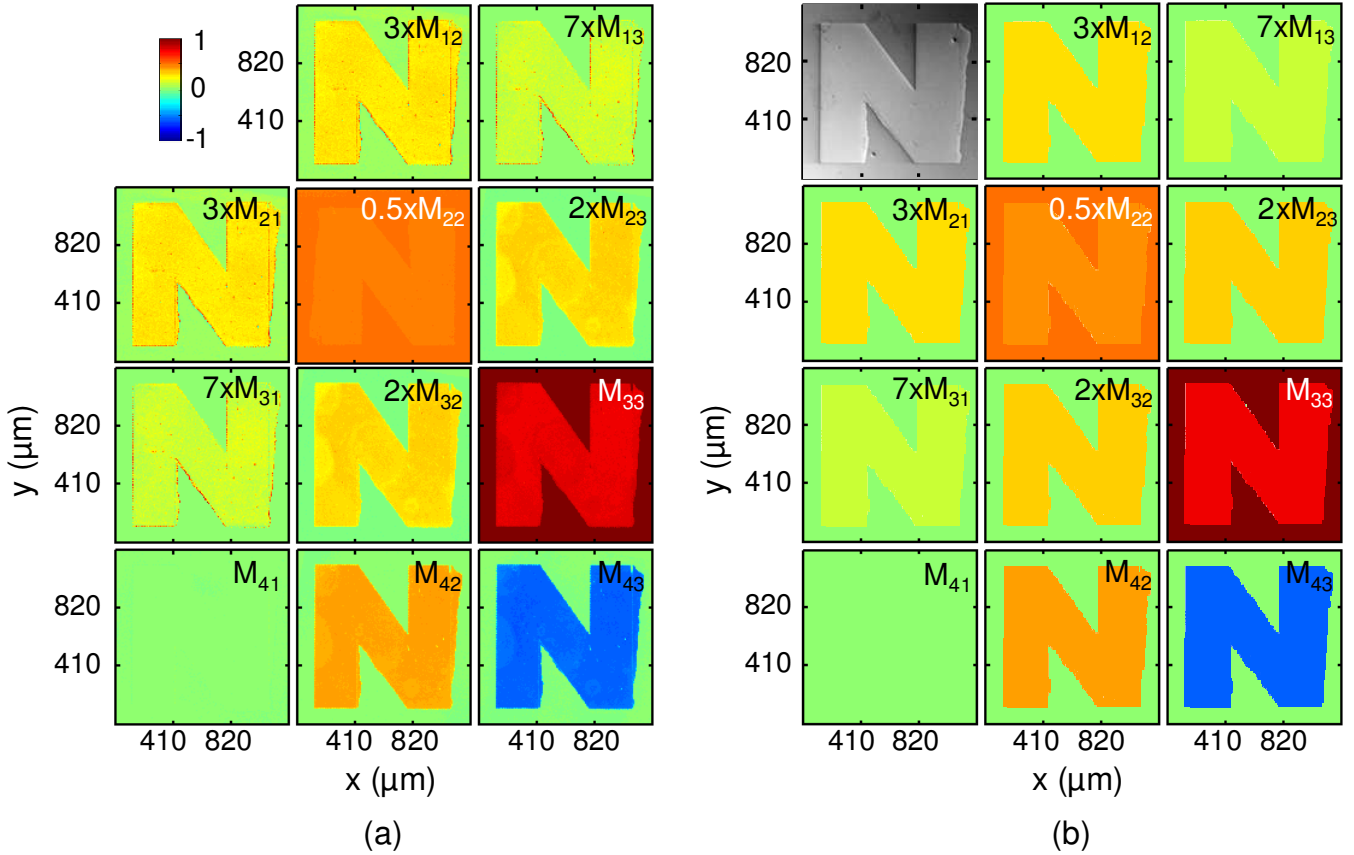}
\caption{(a) ACOM Mueller matrix images of a patterned Si-SCTF sample obtained from Mueller matrix regression of polarized images.  (b) Best-match ellipsometric model calculated ACOM images. The ellipsometric model includes the glass slide and the Si-SCTF. The AB-EMA model is employed to calculate the anisotropic optical properties of the Si-SCTF. The pattern shape is taken from a calibrated SEM images shown as inset in the top row. The lateral extensions of the SCTFs within the N SEM image are exactly in agreement with those in the Mueller matrix images. The Measured Mueller matrix values are in excellent agreement with the model calculations (Field of view: 1.68~mm$\times$1.68~mm; presented area: 1.23~mm$\times$1.23~mm; $\lambda$ = 633~nm; MO: infinity-corrected Nikon CF Plan 5x/0.13na; object image area per pixel $\approx$ 3.3$\times$3.3~$\mu$m$^2$).}
\label{fig:ACMINshape}
\end{figure*}

Anisotropic filters (Sec.~\ref{sec:anisotropicfilter}) with calibrated, patterned areas are prepared by photolithography. A patterned mask for exposure of photoresist is fabricated, and the photo-resist is deposited and exposed prior to GLAD deposition. The GLAD process deposits SCTF using silicon (Si). After Si-SCTF deposition and removal of the photo-resist only Si-SCTFs within the patterned area of the photo-resist layer remain. Figure~\ref{fig:STFpattern} depicts examples of laterally scaled Si-SCTF deposited onto a glass substrate. Scanning electron microscopy (SEM) images are used to obtain lateral dimensions of the patterned Si-SCTF areas, and to reveal their homogeneity. The arrangement of the slanted columns within the Si-SCTF areas is equal to those reported previously. Each ``N shaped'' area is 1~mm$\times$1~mm in lateral dimension. The nominal thickness of the Si-SCTF film is 500~nm. The optical properties of the Si-SCTF are determined from similar Si-SCTF deposited without masks, and characterized by GSE at multiple, oblique angle of incidence measurements as discussed previously.\cite{SchmidtAPL94_2009,SchmidtOL34_2009,SchmidtTSF519_2011,RodenhausenOE20_2012,SchmidtAPL100_2012,KasputisJCC_2013,LiangAPL2013SCTFpolymer,KoenigABC2014,RodenhausenChapter2014SCTF,LiangAPLporousSCTF,WilsonAPL2015graphene,KasputisABM2015} From these investigations, the optical response of the Si-SCTF in normal transmission can be predicted. 

ACOM measurements are performed with the slanting direction (SD) at $\phi$=20$^\circ$ azimuth with respect to $\theta_{\mathrm{A}}=0$ of A. Figure~\ref{fig:ACMINshape} depicts the Mueller matrix images of the patterned SCTF in Fig.~\ref{fig:STFpattern}. The area shown here is 1.23~mm$\times$1.23~mm, and which is centered onto one N shape for convenience. The entire field of view is 1.68~mm $\times$ 1.68~mm. The Si-SCTF converts $p$ polarization into $s$ polarization, and vice versa. This results in non-zero, off-block-diagonal Mueller matrix elements $M_{13}=M_{31},\: M_{23}=M_{32},\: M_{42}$. Figure~\ref{fig:ACMINshape} also depicts calculated images using the AB-EMA model approach discussed in Sect.~\ref{sec:ABEMA}. The model and best-match model parameters used for this calculation were obtained from a separate Si-SCTF grown under the same growth conditions as the sample in Fig.~\ref{fig:ACMINshape}(b) but without patterning, and analyzed by GSE. No actual best-match model analysis is performed for the ACOM images. The model parameters are summarized in the caption of Fig.~\ref{fig:CTABsignal}. The agreement between the experimental and calculated ACOM images is excellent. In the calculated images, the outer boundary dimensions were taken from the SEM images in Fig.~\ref{fig:STFpattern}. We observe a very good agreement among the lateral dimensions between the experimental and calculated Mueller matrix images. Hence, we suggest the use of patterned anisotropic films for quantitative calibration of lateral scales as well as for testing the scales of Mueller matrix values.

\subsection{ACOM detection of small molecule adsorption}
\label{sec:ACOMCTAB}

\begin{figure}[tbp]
\includegraphics[keepaspectratio=true,width=6cm, clip, trim=0 0 0 0 ]{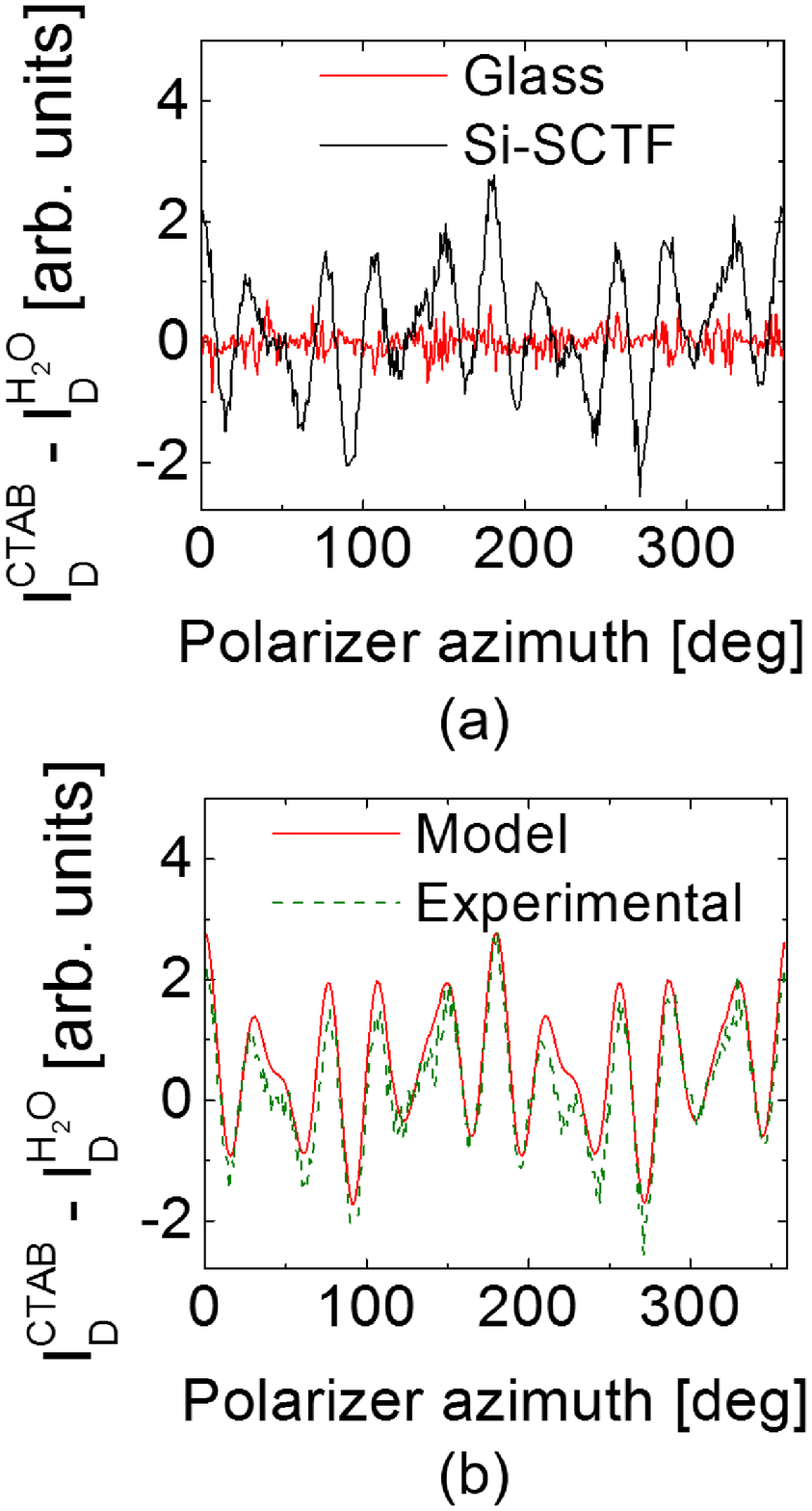}
\caption{Single-group, combined-few-pixel (``binned'') ACOM data $I_{\mathrm{D}}(\theta_{\mathrm{P}},\theta_{\mathrm{C}}=3\theta_{\mathrm{P}})$ shown as differences for a bare glass substrate and glass with Si-SCTF, with and without ultra-thin (2~nm) organic overlayer of cetyltrimethylammonium bromide (CTAB). (a) Measured intensities depicted as differences before and after CTAB adsorption. $\lambda$ = 470~nm. (b) Comparison between the experimental and best-match model calculated data for the CTAB adsorption onto the SCTF. The combined-few-pixel data are taken from ACOM images, which result in the Mueller matrix images shown in Fig.~\ref{fig:CTAB}. Model and model parameters used here and in Fig.~\ref{fig:CTAB}(c): ambient water (dielectric constant $\varepsilon$ = 1.7734), Si-SCTF ($d$ = 500~nm, $f_\mathrm{Si}$ = 23$\%$, $\varepsilon$ = 13.5599 + \textit{i}0.4522), glass ($\varepsilon$ = 2.31) with or without a 2~nm thin film of CTAB ($\varepsilon$ = 2.25) covering conformally the columnar surfaces of the Si-SCTF. The 2~nm coverage corresponds to approximately 10$\%$ CTAB volume fraction within the Si-SCTF. (A conversion chart is given in Ref.~\onlinecite{RodenhausenChapter2014SCTF}, Chapter II.7.)}\label{fig:CTABsignal}
\end{figure}

In this section we demonstrate the detection of ultra-small amounts of organic adsorbates and their lateral distribution within and across the anisotropic filter in the ACOM instrumentation. Surfactants such as cetyltrimethylammonium bromide (CTAB) are useful for nanoparticle synthesis,\cite{WeiASS2005DNAtemplate} and for detergent applications, for example.\cite{vanRuissenJID1998CTABDNA} CTAB adsorption onto Ti-SCTF and flat surfaces was measured recently using a combinatorial quartz crystal microbalance dissipation (QCM-D) and GSE approach by Rodenhausen~\textit{et al.}, where depending on the packing density bi-layers with thickness of about 4~nm form conformal across the surface of either the SCTFs or flat substrates.\cite{RodenhausenTSF2011CTAB,RodenhausenRSI82_2011,RodenhausenChapter2014SCTF} We discuss in this section current limits of detection (sensitivity) for such small organic adsorbates in ACOM. We demonstrate that few femtogram (fg) per square micrometer ($\mu$m$^2$) sensitivity is reached with our current instrumentation, and we compare this limit with typical limits for QCM-D.

Figure~\ref{fig:CTABsignal} depicts calculated and experimental ACOM data for a single group of signal (single group of pixels, or one single pixel) comparing the effect of the adsorption of CTAB onto either a Si-SCTF deposited on glass, or onto a bare glass substrate. We show the original experimental data here, that is, the measured intensity data. In this presentation, the effect of a change in sample properties can be seen in the most pristine form. Note that Mueller matrix data cannot be directly measured in our instrumentation, and are the result of a data model regression analysis. A single group of signal, $I_{\mathrm{D}}(\theta_{\mathrm{P}},\theta_{\mathrm{C}}=3\theta_{\mathrm{P}})$, is depicted versus polarizer azimuth. At each azimuth setting of P, the azimuth orientation of C is three times the setting of P. Hence, the group of signals can be plotted as a single graph. The data shown are differences taken from the group of signal before and after the deposition of CTAB. The ACOM data are obtained from within a liquid cell, which is described further below. 

Figure~\ref{fig:CTABsignal}(a) depicts measured intensities using the ACOM instrumentation combining three pixels into one group of intensity data. The measurements were performed once after the cell was filled with pure water, and a second time after replacement of the fluid with 2.5~mM solution of CTAB. From the 2.5-mM-CTAB solution a homogeneous CTAB thin film forms over the glass surface as well as over the SCTF covering its slanted columns coherently. The two groups of pixels were obtained from a sample region without SCTF, and a region with STCF. The data from the bare glass surface represent the detector noise in this experiment, which means that the few-nm-thick organic overlayer cannot be detected. On the contrary, the modulation detected over the SCTF pixel area follows a distinct pattern, which can be well represented by the ellipsometric model (Fig.~\ref{fig:CTABsignal}(b)), and which permits quantitative evaluation of the amount of CTAB adsorbed within the SCTF. While the presence of CTAB on the bare glass slide cannot be verified at normal incidence using ellipsometric principles, its presence is conveniently measurable by using the anisotropic filter. The signal difference depicted in Figure~\ref{fig:CTABsignal}(a) for glass is what a traditional imaging Mueller matrix microscope would report, where the organic overlayer remains literally invisible. However, the use of the anisotropic filter, and the detection of the polarization modulation clearly reveals the presence of the adsorbate. Figure~\ref{fig:CTABsignal}(a) provides experimental proof of the enhanced contrast obtained in ACOM towards ultra-small amount of an organic specimen.

Figure~\ref{fig:CTABsignal}(b) shows calculated intensity differences for the SCTF upon CTAB adsorption using ellipsometric models.\cite{RodenhausenTSF2011CTAB,RodenhausenRSI82_2011,RodenhausenChapter2014SCTF} The model and model parameters are given in the caption of Fig.~\ref{fig:CTABsignal}. The model calculation follows previously discussed best-match-model ellipsometric approaches for quantification of the adsorption of thick CTAB, using a liquid cell and GSE at oblique angle of incidence, both onto isotropic surfaces and onto SCTF.\cite{RodenhausenTSF2011CTAB,RodenhausenRSI82_2011,RodenhausenChapter2014SCTF} The agreement between experiment and model is excellent. The resulting best-match-model parameter is the fraction, or surface mass density, of CTAB within the combined pixel area, and which is discussed further below. 

Figure~\ref{fig:CTABsignal}(b) also serves as a good example to highlight the importance of choices made for list F[$i,\theta_{\mathrm{P},i},\theta_{\mathrm{C},i}$], that is, the individual polarizer and compensator settings at which images are to be acquired. For the situation discussed in this present example, sufficient settings of P must be included so that positions where maximum changes upon adsorption occur can be detected. For certain applications, in the ACOM instrumentation, the number of required settings in list F[$i,\theta_{\mathrm{P},i},\theta_{\mathrm{C},i}$] may be substantially reduced if the anticipated process is well understood, for example, observation of adsorption or desorption of small amounts of organic or inorganic substances. Reduction of list entries reduces measurement time as well as image analysis computation time.

\begin{figure*}[tbp] \includegraphics[keepaspectratio=true,width=\textwidth, clip, trim=0 0 0 0 ]{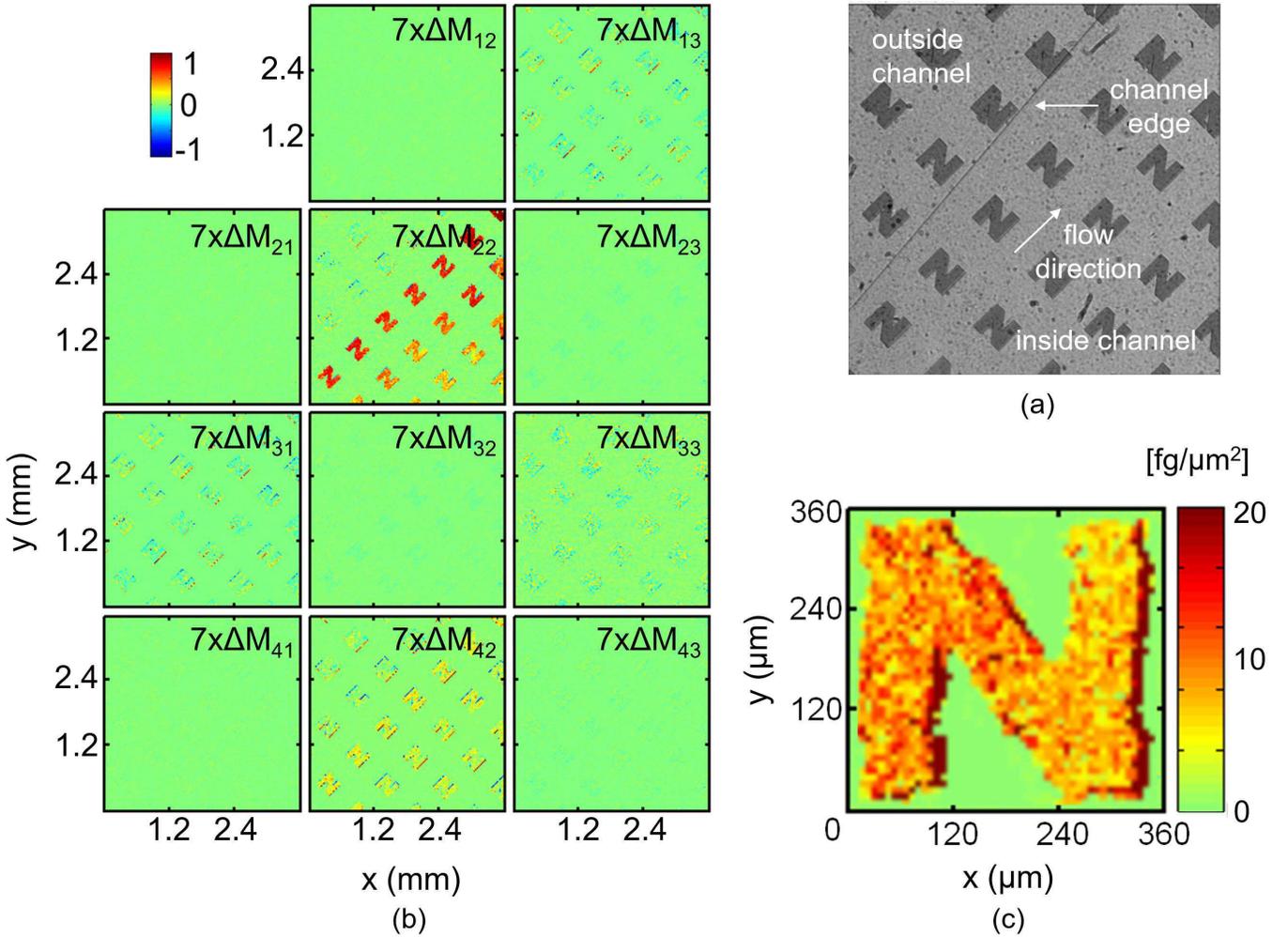}
\caption{(a) Single ACOM instrumentation image, $I_{\mathrm{D}}(\theta_{\mathrm{P}}=0^\circ,\theta_{\mathrm{C}}=0^\circ$), of the transmission flow cell, which comprises a transparent PDMS gasket adhering to the surface of a patterned Si-SCTF on glass (Similar to those shown in Fig.~\ref{fig:STFpattern} with the difference that the N-shaped regions in this experiment are 350~$\mu$m$\times$350~$\mu$m in lateral dimension). The line of the gasket forming the channel is indicated. The flow direction is indicated. The gasket is placed directly onto the Si-SCTF/glass surface. (b) Depicts experimental ACOM Mueller matrix images shown as difference between images obtained within the flow cell after exposure to 2.5~mM-CTAB solution and obtained before with pure water only. The Si-SCTF slanting direction is $45^\circ$ towards analyzer A. The image list F contains $N$ = 360 entries, where $\theta_{\mathrm{C}} = 3 \theta_{\mathrm{P}}$, and $\theta_{\mathrm{P}}$ is moved from 0$^{\circ}$ to 360$^{\circ}$ in steps of 1$^{\circ}$. (c) Shows the CTAB surface mass density obtained as best-match-model parameter from analysis of images shown in Fig.~\ref{fig:CTAB}(b). The noise level in Fig.~\ref{fig:CTAB}(c) is estimated at about 0.33~fg/$\mu$m$^2$, which represents the detection limit of the ACOM instrumentation in this configuration. (Field of view: 3.57~mm$\times$3.57~mm; $\lambda$ = 470~nm; MO: infinity-corrected Olympus Plan N 2x/0.06na; object image area per pixel $\approx$ 7$\times$7~$\mu$m$^2$).}\label{fig:CTAB}
\end{figure*}

For the purpose of imaging the CTAB attachment an~\textit{in-situ} flow cell is constructed. The cell consists of a microscope slide with SCTF, and a transparent gasket forming a flow channel over the SCTF. The slide is patterned with a 500-nm-thick Si-SCTF, analogous to Sec.~\ref{sec:Nimaging} with the difference that the N-shaped regions in this experiment are 350~$\mu$m$\times$350~$\mu$m in lateral dimention. The gasket is made from transparent Polydimethylsiloxane (PDMS). The gasket is prepared in a cylindrical glass mold with depth of 10~mm and cast area diameter 60~mm. A Si-wafer is attached to the bottom of the mold which determines the bottom surface of the gasket with very low roughness. Centered onto the Si-wafer's surface is a 40-$\mu$m-high ridge with lateral dimensions of 15~mm $\times$ 5~mm. Once poured into the mold and polymerized, the PDMS gasket is removed and placed onto the microscope slide forming a microfluidic channel (Fig.~\ref{fig:CTAB}(a)). Two metallic 0.65~mm-diameter stainless-steel syringe needles are inserted through the top of the gasket at the edges of the channel, thereby producing simple inlet and outlet ports to a fluid control device. A programmable syringe pump (New Era Pump Systems, Inc.) pulled solutions through the liquid cell. A HVM-Hamilton valve is used to control the flow. The cell formed thereby possesses an open volume of $\approx$ 3$\mu$l over patterned Si-SCTF inside the microfluidic channel. CTAB is purchased from Sigma-Aldrich, and 18.2 M$\Omega$cm water is obtained from a Barnstead Nanopure water purification system. The flow cell is placed in the ACOM instrumentation with the Si-SCTF slanting direction at 45$^{\circ}$ with respect to A, and the working distance of MO is set to the top of the glass slide with the patterned Si-SCTF. Water is introduced from its respective reservoir to the flow cell at a flow rate of 15$\mu$l/hr. After reaching stable flow, ACOM measurements are performed at $\lambda$ = 470~nm. The flow is then switched from pure water to 2.5~mM CTAB solution at the same flow rate. After a period of at least twice the expected time for the CTAB solution to completely fill the cell, a second ACOM measurement is performed. All data sets are then transformed into Mueller matrix images, and the Mueller matrix images are presented here as differences between those taken after CTAB exposure subtracted by those obtained at pure water flow. These images are shown in Fig.~\ref{fig:CTAB}(b).

The ACOM Mueller matrix element difference images reveal changes in all at locations of the N-shape SCTF areas within the flow channel. It is noted that small amounts of liquid leak under the PDMS gasket, hence small traces of changes in SCTF areas can be detected outside the channel. Furthermore, we also detect a gradient in changes across N shapes towards the center of the channel, which may be due to gradients in flow velocity across the channel. The ACOM difference images can be analyzed by ellipsometric models, in particular the AB-EMA model discussed in Sec.~\ref{sec:ABEMA} is exploited here. Based on a knowledge of the dielectric constant at $\lambda$=470~nm for amorphous Si $\varepsilon_{\mathrm{1}}=13.5599+i0.4522$, pure water $\varepsilon_{\mathrm{2}}=1.7734$, and the organic layer $\varepsilon_{\mathrm{3}}=\varepsilon_{\mathrm{CTAB}}$ = 2.25, Eq.~(\ref{Bruggeman}) can be solved for the volume fraction of the organic layer, $f_{\mathrm{3}}$ = $f_{\mathrm{CTAB}}$. The latter can be used with parameters $d_{\mathrm{SCTF}}$ = 500~nm, and adsorbate density $\rho_{\mathrm{ads}}$ = 0.93~g/ml for calculation of the surface mass density of organic adsorbate onto the SCTF, $\Gamma_{\mathrm{GE}}$.\cite{RodenhausenOE20_2012,RodenhausenChapter2014SCTF} A best-match-model regression is performed for every pixel, and an example for one pixel within the Si-SCTF is shown in Fig.~\ref{fig:CTABsignal}. For every pixel, $f_{\mathrm{CTAB}}$ is determined, and then $f_{\mathrm{CTAB}}$ is plotted versus pixel coordinate. Repeating the described procedure for each pixel, a spatial distribution of $\Gamma_{\mathrm{GE}}$ is evaluated, the result of which is shown in Fig.~\ref{fig:CTAB}(c) for an excerpt of the image area centering on one N shape. 

The amount of attached CTAB detected in this experiment is equivalent to the amount of attachment observed in non-imaging \textit{in-situ} transmission GSE experiments through a similar flow cell with Ti-SCTF described by Rodenhausen~\textit{et al.}\cite{RodenhausenChapter2014SCTF} In their experiment, a commercial spectroscopic ellipsometer was used to determine the anisotropy changes of a Ti-SCTF upon exposure to 2.5~mM CTAB. It was determined that approximately 20$\times 10^{-15}$~g (fg) per $\mu$m$^2$ had attached. However, this experiment was performed by averaging over approximately an area of 3--5~mm in diameter. It is worthwhile to compare the surface mass-per-area detection limits of our current ACOM instrumentation with, for example, QCM-D. The QCM-D technique is commonly used for quantitative determination of the mass-per-area adsorption of small organic molecules onto the QCM-D sensor surface, specifically within liquid environment.\cite{DixonQCMDreview2008,RichterChapter2014QCMDSE} One may discuss the noise level, which must be overcome to register the adsorption event. A definition of a minimum detection signal could then be suggested as the threefold of the signal-to-noise distance required to trigger detection, for example. For QCM-D instrumentation, typical resolution limits for frequency shifts of the QCM-D sensor surface in liquid environments due to mass-per-area attachment is on the order of $\pm$ 0.1 Hz with approximately one order of magnitude better in normal ambient or vacuum.\cite{DixonQCMDreview2008} This leads to sensitivity of few hundreds of pg/cm$^2$. For example, a recent study of aptamer DNA sensor performance obtained sensitivity of 0.1 ng/cm$^2$.\cite{RodenhausenRSI82_2011,RichterChapter2014QCMDSE} However, this needs to be related to the active sensor area in QCM-D. In contemporary equipment, this is a circle with $\approx$ 1 cm$^2$ area. A homogeneous coating over the area is needed for accurate results. Hence, an estimated 500~pg are needed in total for QCM-D detection. With the ACOM we demonstrate attachment of 20 fg/$\mu$m$^2$, and we estimate the current noise limit at 0.33~fg/$\mu$m$^2$ (Fig.~\ref{fig:CTAB}(c)). Hence estimating the current ACOM instrumentation minimum current detection limit at 1~fg/$\mu$m$^2$, and with the current resolution of $\approx$ 7$\times$7$\mu$m$^2$ per pixel (object area imaged onto one single pixel), a total minimum mass of $\approx$ 49~fg is detectable per pixel. This constitutes an improvement of $\approx$ 10,000$\times$ in sensitivity towards mass detection for the ACOM instrumentation over contemporary QCM-D instrumentation. Note that increase in lateral resolution by use of higher-resolving objectives will further increase the sensitivity. QCM-D instrumentation cannot determine the lateral surface mass density distribution across the sensor surface. ACOM instrumentation is capable of spatially resolving the quantity of organic layer adsorbed along the surface of a SCTF. Perhaps more important, very small amounts of organic adsorbates can be detected when the imaged attachment area can be restricted, by microfluidic arrangements for example, to few square micrometers only.

\subsection{ACOM nanoparticle detection}
\label{sec:nanoparticles}

\begin{figure}[tbp]
\includegraphics[keepaspectratio=true,width=8.4cm, clip, trim=0 0 0 0 ]{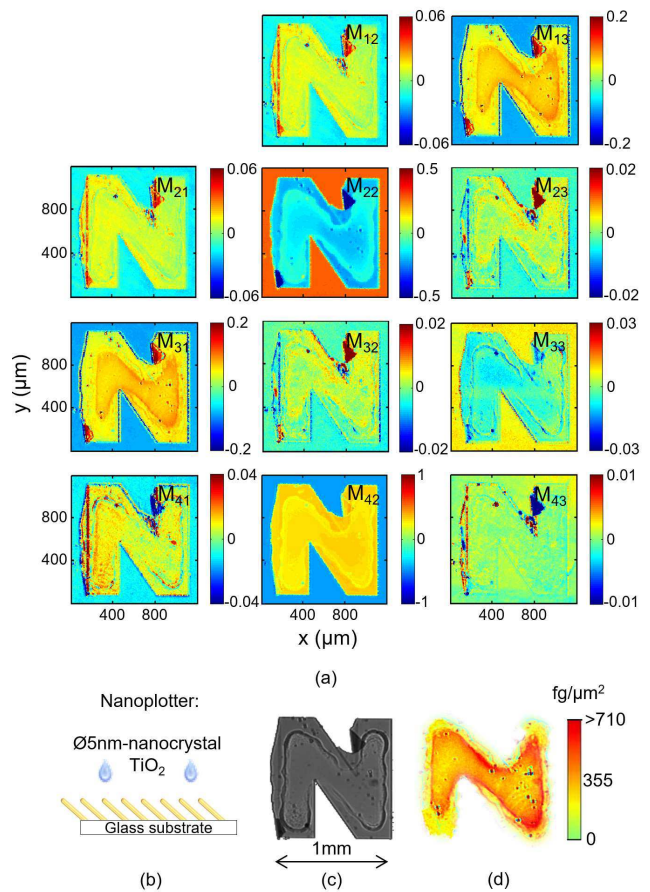}
\caption{(a) ACOM Mueller matrix images of a patterned Si-SCTF after deposition of 177~ng of anatase TiO$_{2}$ nanoparticles. (Note the different color scales for each panel.) The image list F contains $N$ = 360 entries, where $\theta_{\mathrm{C}} = 3 \theta_{\mathrm{P}}$, and $\theta_{\mathrm{P}}$ is moved from 0$^{\circ}$ to 360$^{\circ}$ in steps of 1$^{\circ}$. (b) A Nanoplotter instrumentation is used to dispense the nanoparticles with average diameter of 5~nm in solution on 12 locations along the center line within the N-shaped Si-SCTF. (c) Depicts a single ACOM image $I_{\mathrm{D}}(\theta_{\mathrm{P}}=0^\circ,\theta_{\mathrm{C}}=0^\circ$). (d) Shows the $n$TiO$_{2}$ surface mass density distribution obtained from ellipsometric model analysis of the ACOM Mueller matrix images. The Si-SCTF slanting direction is $45^\circ$ towards analyzer azimuth orientation A. (Field of view: 3.57~mm$\times$3.57~mm; presented area: 1.2~mm$\times$1.2~mm; $\lambda$ = 633~nm; MO: infinity-corrected Olympus Plan N 2x/0.06na; object image area per pixel $\approx$ 7$\times$7~$\mu$m$^2$).}\label{fig:Particles}
\end{figure}

Titanium dioxide nanoparticles ($n$TiO$_{2}$) are currently the most extensively manufactured engineered nanomaterials.\cite{NakataJPP2012,WangNR2004,ZuckerCPA2010} Soil contamination is a growing concern, and thus the detection of nanoparticles is of contemporary interest.\cite{KananizadehJHM2016} Here we present detection of $n$TiO$_{2}$ using the ACOM instrumentation, where $n$TiO$_{2}$ are infiltrated into the anisotropic SCTF. Anatase $n$TiO$_{2}$ stabilized by polyacrylate sodium are purchased from Sciventions, Inc. The average particle diameter is 5~nm. A nanoplotter instrumentation (Nanoplotter 2.0, GeSIM) is used for accurate and controlled infiltration of $n$TiO$_{2}$ into patterned SCTF. The patterned Si-SCTF are prepared as in the previous paragraphs. The dimension of the N shape is 1~mm $\times$ 1~mm. A solution with concentration of 1.5~mg/$\mu$m$^3$ of $n$TiO$_{2}$ is used for printing. 120 drops with individual volume of 1 pm$^3$ are dispensed onto 12 spots (10 drops each) along the center line of a patterned Si-SCTF sample (Fig.~\ref{fig:Particles}(b,c)). The total mass of $n$TiO$_{2}$ dispensed thereby is 177 ng. ACOM measurement are performed 30~min after the solution is dispensed, and the solvent is evaporized. The azimuth orientation of the Si-SCTF sample is set to $45^\circ$ with respect to A. Images are taken at $\lambda$ = 633~nm, and shown in Fig.~\ref{fig:Particles}(a).  

A linearization approach is implemented for a simplified ellipsometric model analysis of the ACOM images. For small changes of volume fraction of adsorbed nanoparticles within a Si-SCTF the off-diagonal-block Mueller matrix elements change linearly, which can be verified by AB-EMA model calculations. Important in this evaluation is the fact that areas in the N-shape can be identified which are unaffected by $n$TiO$_{2}$, and which can be used as the zero-point for the linear extrapolation (regions of no mass attachment). This is possible in this experiment because of the ``coffee-mug-stain-effect'' seen in Fig.~\ref{fig:Particles}(c), where the solution of $n$TiO$_{2}$, which is nanoplotted into the SCTF does not disperse throughout the entire N-shape area due to fast evaporation of the solvent. This process can be well controlled by choice of drop size, solvent, and nanoplotter repetition time. Because the exact optical constants for $n$TiO$_{2}$ are unknown, we use the unknown but assumed linear relationship between variations in Mueller matrix elements $M_{23}=M_{32}$ and $n$TiO$_{2}$ fraction across the N shape. In order to determine the linear scale factor, and because the total mass of $n$TiO$_{2}$ within the N shape as well as the N-shape area are known, we determine the average over all changes in $M_{23,32}$ within the N shape. This value determines exactly the value at which half of the total mass per surface area is located. Hence, the color image scale bar code for $M_{23}$ can be substituted, and the scale value of $M_{23,32}^{av}$ then equals exactly one half of the total mass per total area of the N shape. Then, the same color scale renders the spatial distribution of the surface mass distribution of the $n$TiO$_{2}$ across the N shape. This result is shown in the image in Fig.~\ref{fig:Particles}(d).

\subsection{ACOM observation of lipophilic test dye transport in SCTF}
\label{sec:ACOMDYE}

SCTF as anisotropic filters in ACOM can be used for both imaging and in chemical separations through the use of ultra-thin layer chromatography (UTLC).\cite{HauckJCS40_2002,MorlockAC82_2010,BezuidenhoutLC11_2011,JimAC82_2010,OkoJCA1249_2012,OkoJCA1218_2011}  In this case, the surface and structure of the SCTF is exploited as a stationary phase\cite{PooleJChrom2011,BezuidenhoutLC11_2011} that could be used for the retention and resolution of applied target analytes.\cite{PooleBook} The combination of this chromatographic technique with ACOM permits for the simultaneous separation and detection of targets during their separation on the SCTF support. Detection in this case is based on the imaging of the anisotropy changes within the anisotropic filter.

\begin{figure}[]
\includegraphics[keepaspectratio=true,width=7.5cm, clip, trim=0 0 0 0]{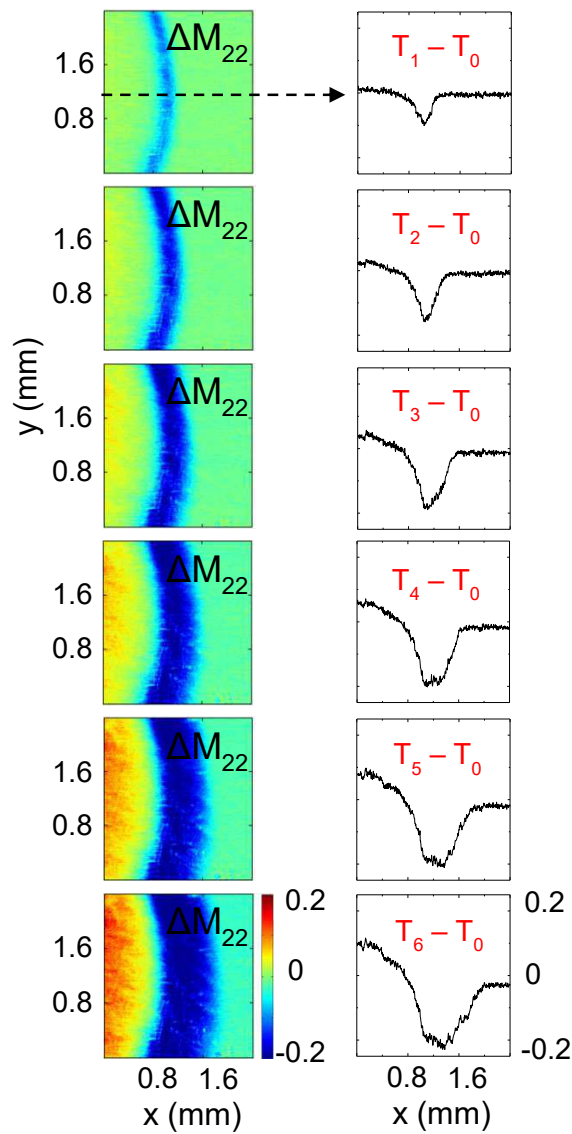}
\caption{ACOM Mueller matrix differences $\Delta$M$_{22}$ (left column: surface images; right column: line cross sections) of a separation of lipophylic dyes into overlapping bands of a patterned SiO$_2$-SCTF. Data are plotted as differences between time T$_0$ and data taken at subsequent time intervals of 45~sec. The initial dye location (left side of each graph) continuously moves towards the right while separating into individual bands. The image list F contains $N$ = 120 entries, where $\theta_{\mathrm{C}} = 3 \theta_{\mathrm{P}}$, and $\theta_{\mathrm{P}}$ is moved from 0$^{\circ}$ to 360$^{\circ}$ in steps of 3$^{\circ}$. (Field of view: 2.4~mm$\times$2.4~mm; $\lambda$ = 633~nm; object image area per pixel $\approx$ 4.7$\times$4.7~$\mu$m$^2$).}
\label{fig:ACOMDye}
\end{figure}

Initial studies with this system were conducted by using a series of colored and lipophilic dyes as model analytes (i.e., test dye mixture III, CAMAG, CH-4132 Muttenz 1, Switzerland). The SCTF was prepared on a glass substrate by using GLAD deposition to produce SiO$_2$ columns with the lengths of 2.5-3.0~$\mu$m. These columns were then coated with an ultra-thin layer of alumina that was deposited by using ALD, as described in Ref.~\onlinecite{SchmidtAPL100_2012}. The mobile phase that was used in these studies was a 4:3 mixture of toluene and n-hexane (purchased from Sigma Aldrich), which is known to allow a separation for many of the same dye components on a traditional alumina support for thin-layer chromatography.\cite{ShermaJESHB2009,ShermaJAOACI2010}  Approximately 90~nL of the test dye mixture was applied as a spot to the SCTF serving as a UTLC plate and allowed to dry at room temperature. This plate was then placed into an enclosed glass chamber, placed into contact with the mobile phase through wick flow and allowed to develop for approximately 30~min.  Within that time period, movement of the dyes was confirmed through visual inspection while movement of the dyes was also recorded through the ACOM imaging system (Fig.~\ref{fig:ACOMDye}). Mueller matrix data suggested that within 6~minutes the dye was completely transferred across the SCTF. Figure~\ref{fig:ACOMDye} depicts the time evolution of the dye transort and separation, visualized here by difference data between data taken at the beginning of the separation and data taken at later points in time. In the lateral cross sections, positive data ($\Delta$M$_{22}$) indicate SCTF regions where dye is removed, while negative data indicate regions where dye is entering. As time progresses, one can identify the initial nearly Gaussian-shaped transport front separating into multiple fronts corresponding to the test dye mixture spotted onto the SCTF. A separation of the dyes into overlapping bands was observed within a travel distance of only 5-6~mm, indicating that both chromatographic separations and imaging are possible with ACOM. This may lead ACOM towards a new approach in UTLC: imaging chromatography.

\subsection{ACOM imaging of living mouse fibroblast cells}
\label{sec:cells}

\begin{figure}[tbp]
\includegraphics[keepaspectratio=true,width=8.2cm, clip, trim=0 0 0 0]{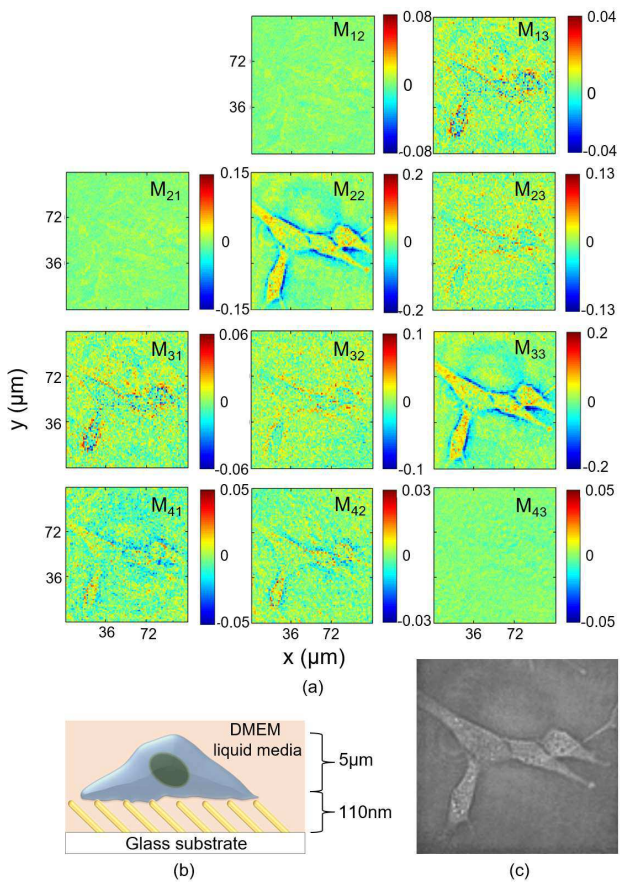}
\caption{(a) ACOM Mueller matrix images of mouse fibroblasts cultured onto 110-nm-thick Ti-SCTF on microscope glass slides. The Mueller matrix values within every block of 5$\times$5 pixels are averaged (binning). The scale of each image is rescaled by subtracting the mean, taken for a Ti-SCTF 25$\times$25px area away from the cells, M$^{av}_{ij}$ from every image pixel point. The Ti-SCTF slanting direction is $45^\circ$ towards A. The image list F contains $N$ = 120 entries, where $\theta_{\mathrm{C}} = 3 \theta_{\mathrm{P}}$, and $\theta_{\mathrm{P}}$ is moved from 0$^{\circ}$ to 360$^{\circ}$ in steps of 3$^{\circ}$. (b) Depicts a graphical representation of a cell situated within Dulbecco’s Modified Eagles Media (DMEM) on top of the Ti-SCTF (not to scale). (c) Shows a single ACOM image $I_{\mathrm{D}}(\theta_{\mathrm{P}}=0^\circ,\theta_{\mathrm{C}}=0^\circ$). (Field of view: 108.94~$\mu$m$\times$108.94~$\mu$m; $\lambda$ = 633~nm; MO: infinity-corrected Olympus ULWD MSPlan 80$\times$/0.75na; object image area per binned 5$\times$5 pixels $\approx$ 1.1$\times$1.1~$\mu$m$^2$).}
\label{fig:cells}
\end{figure}

In this application, SCTF fabricated from titanium onto glass microscope slides are used for image living cells by the ACOM instrumentation. To date, fluorescent microscopy techniques, such as confocal microscopy, provide ample details on cell and subcellular components, such as fluorescently-labeled cellular features, organelles, or molecular factors (e.g., proteins or nucleic acids), but also require destructive manipulation of the cell by means of staining and fixing procedures. The approach presented here using the ACOM instrumentation permits an alternative modality for noninvasive probing of cellular features and cell-material interactions. This approach may be useful for evaluating biomaterial interfaces (e.g., in terms of biomolecule adsorption or cellular adhesion), as well as cellular features (podia or intracellular features), which could have applications in drug and gene delivery, sensors and diagnostics, medical devices and tissue engineering. In contrast to traditional microscopy techniques, where cells are commonly imaged on flat substrates, in the ACOM instrumentation, the nanostructured, anisotropic filter enhances the contrast to image cells. The cells may either be attached to the SCTF or in its close vicinity. The SCTF itself may also provide extracellular cues to the cells,\cite{KasputisABM2015} which could be analyzed through ACOM. 

The Ti-SCTF are prepared as described previously, except these are not patterned. The thickness of the Ti-SCTF is 110~nm. The Ti-SCTF is sterilized by immersing in 200-proof ethanol, followed by transferring the sample to a sterile laminar flow hood to air-dry. Then, the sample was rinsed twice in 1X phosphate buffered saline (PBS), followed by the application of a 10~$\mu$g/ml solution of fibronectin protein (FN) dissolved in PBS to coat the Ti-SCTF sample with a layer of FN extracellular matrix protein to enhance cell adhesion.\cite{KasputisABM2015} After 90~min in FN solution, the sample is rinsed with 1X PBS and NIH/3T3 mouse fibroblasts (cultured in Dulbecco’s Modified Eagles Media (DMEM), supplemented with 10\% fetal calf serum and 1\% penicillin/streptomycin) are then seeded at a concentration of 50,000 cells/ml and cultured in an incubator for 24~h at 37$^{\circ}$C in 5\% CO$_{2}$ atmosphere. On the following day, the sample is transferred to a 10~cm$^2$ Petri dish containing warm media and placed onto the sample stage of the ACOM instrumentation. The ACOM Mueller matrix images are shown in Fig.~\ref{fig:cells}(a). The Mueller matrix values within every block of 5$\times$5 pixels are averaged (binning). Figure~\ref{fig:cells}(b) shows a schematic drawing of a cell located ontop of the Ti-SCTF. A single ACOM image is also shown at $I_{\mathrm{D}}(\theta_{\mathrm{P}}=0^\circ,\theta_{\mathrm{C}}=0^\circ$) in Fig.~\ref{fig:cells}(c). The Mueller matrix images reveal the location and distribution of the cell across the surface of the anisotropic filter. Ellipsometric model analysis methods will be developed in order to differentiate between changes observed due to interaction of the cell with the Ti-SCTF, and in order to quantify, for example, surface mass density and partial infiltration (e.g., focal adhesion) of the cell within the Ti-SCTF. We also expect that the cell may affect the local orientation of the slanting angle of the columnar nanostructures due to interaction of the cell with the surface of the substrate. While this is the topic of future work, we believe that the images presented here demonstrate an alternative imaging modality for cell studies. The ACOM instrumentation also offers an interesting approach to study protein and cellular interactions on nanoscale features.

\section{Summary}

We described a setup to obtain polarized microscopic images of specimen placed within the object plane of a traditional microscopy setup. We have augmented linear polarizers and one compensator to determine the Mueller matrix elements of the object plane using ellipsometric principles. In particular, the novelty of our instrumentation consists of the use of an anisotropic filter, which is placed within the object plane. The anisoptropic filter used here consists of highly-ordered nanostructured thin films prepared by glancing angle deposition, slanted columnar thin films. We described theoretical model approaches to calculate the effect of the anisotropic filter onto the formation of images. We presented approaches for calibration and for operation of the instrumentation. We demonstrated the instrumentation and its performance by measuring the amount of attached mass per surface area for ultra-thin, organic overlayers within the anisotropic filter, by measuring the distribution of nanoparticles, by observing the transport and separation of test dyes and by observation of living cells cultured onto the anisotropic filter. We believe that the approach described in this work will become a useful technique for the study of interaction and presence of organic and inorganic substances with anisotropic and nanostructured substrates.

\begin{acknowledgements}
This work was supported in part by the National Science Foundation (NSF) through the Center for Nanohybrid Functional Materials (EPS-1004094), the Nebraska Materials Research Science and Engineering Center (DMR 1420645), CAREER award CBET 1254415, and awards CMMI 1337856, CHE 1309806, EAR 1521428. The authors further acknowledge grant support by the University of Nebraska-Lincoln, the J.~A.~Woollam Co., Inc., and the J.~A.~Woollam Foundation. The research was performed in part in the Nebraska Nanoscale Facility: National Nanotechnology Coordinated Infrastructure and the Nebraska Center for Materials and Nanoscience, which are supported by the National Science Foundation under Award ECCS: 1542182, and the Nebraska Research Initiative.
\end{acknowledgements}

\bibliography{CompleteLibrary_dar}

\begin{thebibliography}{125}
\bibitem[{foo({\natexlab{b}})}]{footnote3RSIACOM2016}
\bibinfo{journal}{Dark-field illumination modes are often used, for example in
  Zernike and petrographic microscopes, enhancing specimen contrast caused by
  scattering. In principle, ACOM can also operate in dark-field illumination
  mode, which is however not discussed here. Properties and merits of
  illuminating configurations enhancing specimen contrast due to scattering
  remain to be investigated in ACOM.}  (????{\natexlab{b}}).
\bibitem[{foo({\natexlab{d}})}]{footnote4RSIACOM2016}
\bibinfo{journal}{The anisotropic filter may possess one out of, or any
  combination therefrom, linear birefringence (LB), circular birefringence,
  linear dichroism (LD), and circular dichroism (CD). For example, a particular
  ACOM instrument may operate as CD-ACOM, if the anisotropy of the filter is
  purely CD.}  (????{\natexlab{d}}).
\bibitem[{foo({\natexlab{g}})}]{footnote7RSIACOM2016}
\bibinfo{journal}{This absence is of no further concern at this point because
  the Mueller matrix is symmetric for specimens investigated in this work. For
  full expansion of the concept described here, for example to a specimen with
  non-reciprocal properties composed of chiral materials or magnetized
  materials, a second compensator must be incorporated in order to access the
  full Mueller matrix.}  (????{\natexlab{g}}).
\end{thebibliography}

\end{document}